\documentclass[10pt]{ietbook}
\usepackage{algpseudocode}
\usepackage{multirow}
\usepackage{longtable}
\usepackage{lstautogobble}
\usepackage[linesnumbered,ruled,vlined]{algorithm2e}
\usepackage{subfigure}

\usepackage{mathtools}

\usepackage{booktabs}
\usepackage{listings}
\usepackage[colorlinks,pdfstartview=FitH,citecolor=blue,linkcolor=blue,urlcolor=blue]{hyperref}
\usepackage{color}
\usepackage{xcolor}
\usepackage{pythonhighlight}
\definecolor{codegreen}{rgb}{0,0.6,0}
\definecolor{codegray}{rgb}{0.5,0.5,0.5}
\definecolor{codepurple}{rgb}{0.58,0,0.82}
\definecolor{backcolour}{rgb}{0.95,0.95,0.92}

\newcommand\blfootnote[1]{%
  \begingroup
  \renewcommand\thefootnote{}\footnote{#1}%
  \addtocounter{footnote}{-1}%
  \endgroup
}

\lstdefinestyle{mystyle}{
    backgroundcolor=\color{backcolour},   
    commentstyle=\color{codegreen},
    keywordstyle=\color{magenta},
    numberstyle=\tiny\color{codegray},
    stringstyle=\color{codepurple},
    basicstyle=\ttfamily\footnotesize,
    breakatwhitespace=false,         
    breaklines=true,                 
    captionpos=b,                    
    keepspaces=true,                 
    numbers=left,                    
    numbersep=5pt,                  
    showspaces=false,                
    showstringspaces=false,
    showtabs=false,                  
    tabsize=2
}

\begin{document}
\sloppy
\rhbooktitle{Managing Internet of Things Applications across Edge and Cloud Data Centres}

\markboth{Managing Internet of Things Applications across Edge and Cloud Data Centres}{Resource Management in Edge and Fog Computing using FogBus2 Framework}
\cauthor{Mohammad Goudarzi\blfootnote{M. Goudarzi, Q. Deng, and R. Buyya are with The Cloud Computing and Distributed Systems (CLOUDS) Laboratory, School of Computing and Information Systems, The University of Melbourne, Australia.}, Qifan Deng, and Rajkumar Buyya}

\setcounter{chapter}{18}
\chapter{Resource Management in Edge and Fog Computing using FogBus2 Framework}

\section{Introduction}
The rapid advancements of hardware, software, and communication technologies enable the Internet of Things (IoT) to offer a wide variety of intelligent solutions in every single aspect of our lives. Therefore, IoT-enabled systems such as smart healthcare, transportation, agriculture, and entertainment, just to mention a few, have been attracting ever-increasing attention in academia and industry. IoT applications generate a massive amount of data which requires processing and storage, while IoT devices often lack sufficient processing and storage resources. Cloud computing offers infrastructure, platform, and software services for IoT-enabled systems, through which IoT applications can process, store, and analyze their generated data in surrogate Cloud Servers (CSs) \cite{hu2017survey,goudarzi2016efficient}. There are different Cloud Service Providers (CSPs) with a wide variety of services, where each CSP provides a particular set of services such as computing, database, and data analysis in an optimized way. Hence, no CSP can satisfy the full functional requirements of different IoT applications in an optimized manner \cite{li2013applications}. As a result, each IoT application can be particularly serviced by a specific CSP or simultaneously by different CSPs, which is often called hybrid cloud computing \cite{li2013applications}. Although hybrid cloud computing platform provides IoT devices with unlimited and diverse computing and storage resources, CSs are residing multi-hops away from IoT devices, which incurs high propagation and queuing latency. Thus, CSs cannot solely provide the best possible services for latency-critical and real-time IoT applications (e.g., intelligent transportation, smart healthcare, emergency, and real-time control systems) \cite{schulz2017latency,goudarzi2020application}. Besides, forwarding the huge amount of data generated by distributed IoT devices to CSs for processing and storage may overload the CSs \cite{mahmud2019modelling}. To overcome these issues, edge and fog computing has emerged as a novel distributed computing paradigm.
\par
\begin{figure}[t]
    \centering
    \includegraphics[width=\linewidth,height=6cm]{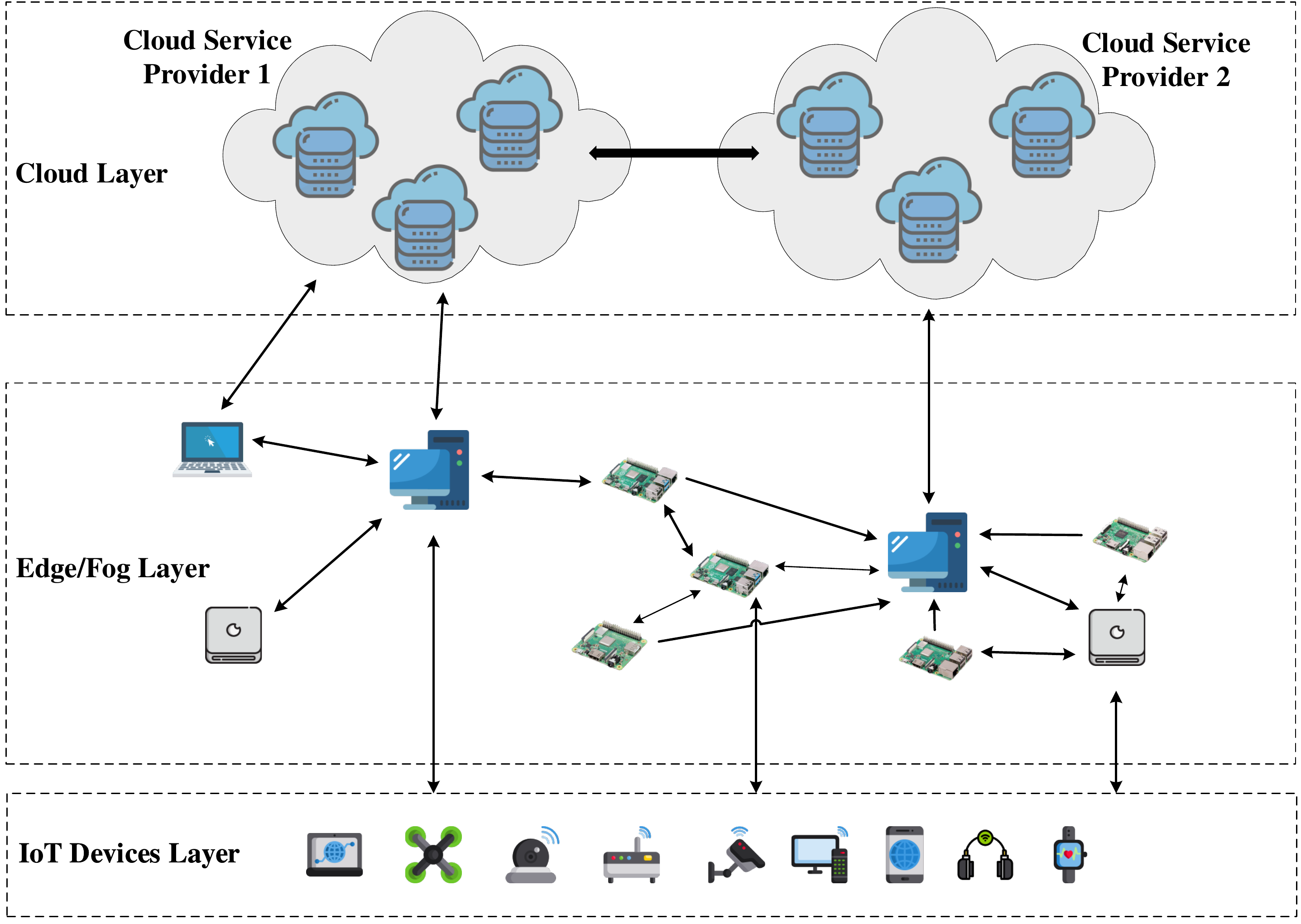}
    \caption{Heterogeneous Computing Environment containing Multiple Cloud Servers, Edge/Fog Servers, and IoT Devices}
    \label{fig:heterogeneousComputingEnvironment}
\end{figure}

In edge and fog computing environments, the geographically distributed heterogeneous Edge Servers (ESs) (e.g., access points, smartphones, Raspberry-Pis), situated in the vicinity of IoT devices, can be used for processing and storage of IoT devices' data. These ESs can be accessed with lower latency, which makes them a potential candidate for latency-critical IoT applications, and reduce the traffic of the network's backbone \cite{goudarzi2019fog}. However, the computing and storage resources of ESs are limited compared to CSs, so that they cannot efficiently execute computation-intensive tasks. Therefore, to satisfy the resource and Quality of Service (QoS) requirements of diverse IoT-enabled systems, a seamlessly integrated computing environment with heterogeneous edge/fog and different cloud infrastructures is required, as depicted in Figure \ref{fig:heterogeneousComputingEnvironment}.  

\par
The computing and storage resources in such an integrated environment are highly heterogeneous in terms of their architecture, processing speed, RAM capacity, communication protocols, access bandwidth, and latency, just to mention a few. Furthermore, there are a wide variety of IoT-enabled systems with various QoS and resource requirements. Accordingly, to satisfy the requirements of IoT applications in such an integrated environment, scheduling and resource management techniques are required to dynamically place incoming requests of IoT applications on appropriate servers for processing and storage \cite{merlino2019enabling}. In order to develop, test, deploy, and analyze different IoT applications and scheduling and resource management techniques in real-world scenarios, lightweight and easy-to-use frameworks are required for both researchers and developers. There are some existing frameworks for integrating IoT-enabled systems with edge and fog computing such as \cite{tuli2019fogbus,yousefpour2019fogplan,merlino2019enabling,nguyen2019market, bellavista2017feasibility,ferrer2019ad,noor2019iotdoc}. However, they only focus on one aspect of IoT-enabled systems in edge and fog computing, such as scheduling, implementation of a new type of IoT application, or resource discovery. In this chapter, we provide a tutorial on FogBus2 framework \cite{deng2021fogbus2} which offers IoT developers a suite of containerized IoT applications, scheduling and scalability mechanisms, and different resource management policies in an integrated environment, consisting of multi-cloud service providers, edge and fog computing servers, and IoT devices. Furthermore, we extend this framework by new resource management techniques such as a new scheduling policy. In addition, new types of IoT applications, either real-time or non-real-time, are implemented and integrated with the FogBus2 framework.

\par
The rest of the chapter is organized as follows: We start with discussion on the FogBus2 framework, its main components, and respective communication protocols. Next, we describe how to install and run the current functionalities of FogBus2, considering different IoT applications. Finally, we provide a guideline presenting how to develop and integrate new IoT applications and new policies into the FogBus2.

\section{FogBus2 Framework}
FogBus2 \cite{deng2021fogbus2} is a new container-based framework based on docker containers, developed in Python. To enable the integration of various IoT application scenarios in highly heterogeneous computing environments, FogBus2's components can be simultaneously executed on one or multiple distributed servers in any computing layer. This feature significantly helps researchers and developers in the development and testing phases because they can develop, test, and debug their desired IoT applications, scheduling, and resource management policies on one or a small number of servers. Furthermore, in the deployment phase, they can run and test their IoT applications, scheduling, and resource management techniques on an unlimited number of servers.

\subsection{Main Components}
\begin{figure}[t]
    \centering
    \includegraphics[width=\linewidth,height=7cm]{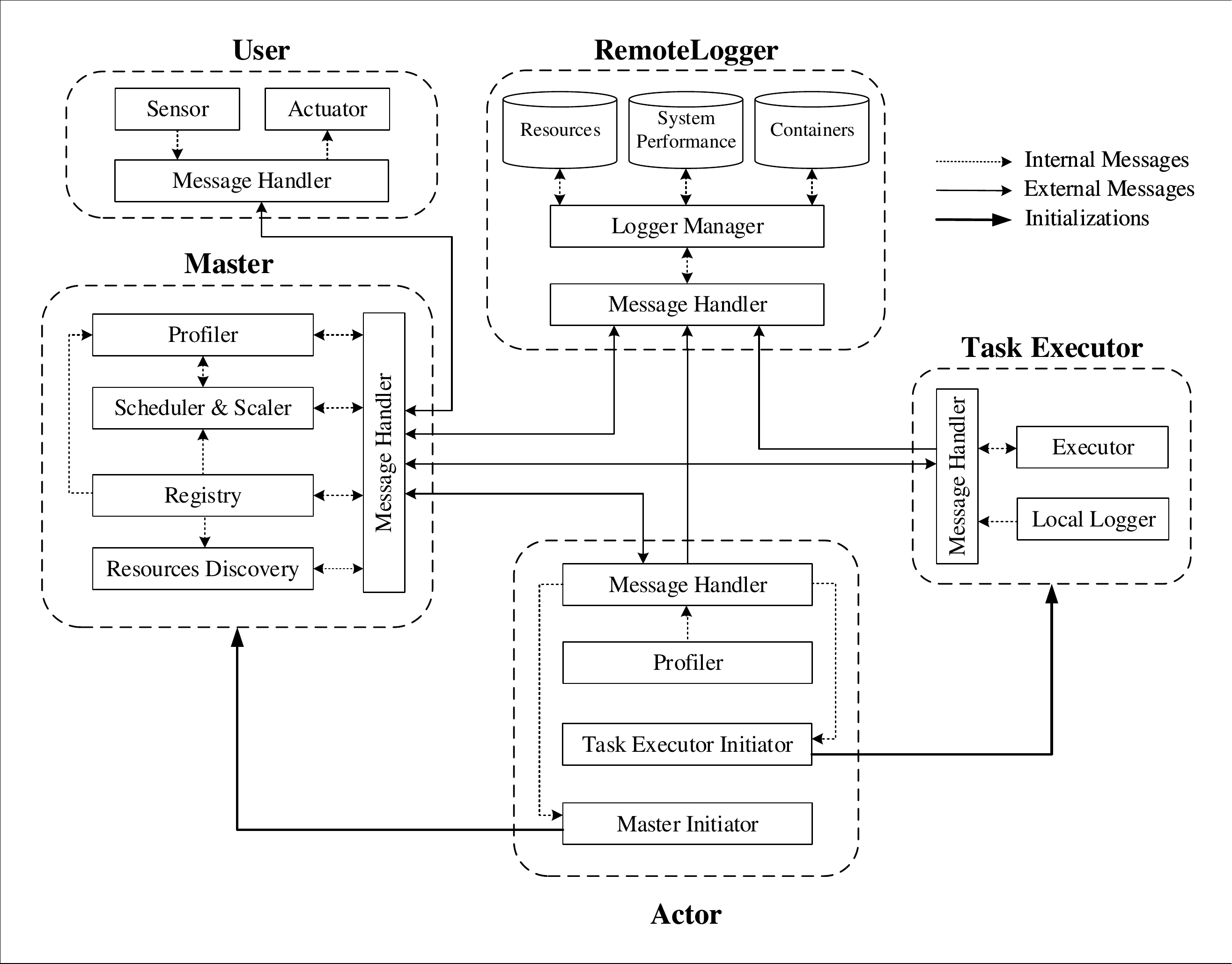}
    \caption{FogBus2 Main Components, Sub-components, and their Interactions}
    \label{fig:Fogbus2components}
    \vspace{-3mm}
\end{figure}

FogBus2 consists of five containerized components, namely \textit{User}, \textit{Master}, \textit{Actor}, \textit{Task Executor}, and \textit{Remote Logger}. Among these components, the \textit{User} should run on IoT devices or any servers that directly interact with users' sensory or input data. The rest of the components can run on any servers with sufficient resources. Each of these containerized components contains several sub-components (sub-C) with specific functionalities. Figure \ref{fig:Fogbus2components} presents FogBus2's main components and their respective sub-Cs. Since the components of the FogBus2 can run on geographically distributed servers, a \textit{message handler} Sub-C is embedded in each component to handle sending and receiving of messages. In what follows, we briefly describe the main functionalities and sub-Cs of each component.

\begin{itemize}

\item{\textbf{User}:} 
This component controls the IoT device's requests for surrogate resources and contains two main sub-Cs, namely \textit{sensor} and \textit{actuator}, alongside with \textit{message handler}. The \textit{sensor} is responsible for capturing and management of raw sensory data and configuring sensing intervals based on IoT application scenarios. Besides, the \textit{actuator}'s main function is collecting the incoming processed data and executing a respective action. The \textit{actuator} can be configured by its users to perform real-time actions based on incoming processed data or periodic actions based on a batch of processed data. Researchers and developers can configure the \textit{sensor} and \textit{actuator} to implement different application scenarios.

\item{\textbf{Remote Logger}:}
The main functionality of this component is to collect and store the contextual information of servers, IoT devices, IoT applications, and networking. It contains the \textit{logger manager} sub-C that can connect to different databases, receives logs of other components, and stores logs in persistent storage. By default, the \textit{Remote Logger} connects to databases to store logs, which is easier to manage and maintain. However, logs can be stored in files as well.    

\item{\textbf{Master}:}
In a real-world computing environment, one or multiple \textit{Master} components may exist. This component contains four main sub-Cs, called \textit{profiler}, \textit{scheduler $\&$ scaler}, \textit{registry}, and \textit{resource discovery}, alongside with the \textit{message handler}. When the \textit{Master} starts, the \textit{resource discovery} sub-C periodically search the network to find available \textit{Remote Logger}, \textit{Master}, and \textit{Actor} components in the network. If new components can be found in the network, the \textit{resource discovery} advertise itself to those components, so that they can send a request and register themselves in this \textit{Master}. If the \textit{Master} receives any requests for registration or placement requests from IoT devices (i.e., \textit{User} components), the \textit{registry} sub-C will be called. This sub-C records the information of IoT devices and other components and assigns them a unique identifier. Besides, when the incoming message is a placement request from \textit{User} components, it initiates the \textit{$scheduler\;\&\; scaler$} sub-C. The \textit{$scheduler\;\&\; scaler$} sub-C receives the placement request from the \textit{registry} sub-C, the contextual profiling information of all available servers, and networking information from the \textit{profiler} sub-C. Next, if it has enough resources to run the scheduling policy and its placement queue is not very large (configurable queue size), it runs one of the scheduling policies implemented in the FogBus2 framework to assign tasks/containers of the IoT application on different servers for the execution. According to the outcome of scheduling policy, the \textit{Master} component forwards required information to the selected \textit{Actors} to execute tasks/containers of the IoT application. Currently, three scheduling policies are embedded in the FogBus2 framework, namely Non-dominated Sorting Genetic Algorithm 2 (\textit{NSGA2}) \cite{deb2002fast}, Non-dominated Sorting Genetic Algorithm 3 (\textit{NSGA3}) \cite{deb2013evolutionary}, and Optimized History-based Non-dominated Sorting Genetic Algorithm (\textit{OHNSGA}) \cite{deng2021fogbus2}. If due to any reason the \textit{Master} component cannot run its scheduling policy, it runs the \textit{scalability} mechanism to forward the placement request to other available \textit{Master} components, or it initiates a new \textit{Master} component on of the available servers. In the rest of this chapter, we describe how to use current scheduling and scalability policies. Furthermore, we also present how to implement new scheduling and scalability policies and integrate them into the FogBus2 framework.                       
 
\item{\textbf{Actor}:}
The main responsibility of this component is to start different \textit{Task Executor} components on the server on which it is running. To illustrate, available surrogate servers in the environment should run \textit{Actor} component. Then, these \textit{Actor} components will be automatically discovered and registered by one or several \textit{Master} components in the environment. The \textit{Actor} component profiles the hardware and networking condition of the server on which it is running using the \textit{profiler} sub-C. Besides, when a \textit{Master} component assigns a task of an IoT application to an \textit{Actor} for the execution, it calls the \textit{task executor initiator} sub-C which initiates different \textit{Task Executor} components on the server according to different IoT applications. This sub-C also defines the destination to which the result of each \textit{Task Executor} should be forwarded based on the dependency model of the IoT application. Finally, in order to scale \textit{Master} components in the environment, each \textit{Actor} is embedded with a \textit{master initiator} sub-C. When an \textit{Actor} receives a scaling message from one of the available \textit{Master} components in the environment, the \textit{master initiator} sub-C will be called. This sub-C starts a \textit{Master} component on the server, which can independently serve incoming IoT application requests. In addition, it can be seen that each server simultaneously can run different components (e.g., \textit{Master}, \textit{Actor}, \textit{Task Executor}, etc) and play different roles.

\item{\textbf{Task Executor}:}
IoT applications can be represented as a set of dependent or independent tasks or services. In the rest of this chapter, tasks and services are used interchangeably.  In the dependent model, the execution of tasks has constraints and each task can be executed when its predecessor tasks are properly executed. In FogBus2, each \textit{Task Executor} component is responsible for the execution of specific task; i.e., each task or service can be containerized as a \textit{Task Executor}. To illustrate, an IoT application with three decoupled tasks should have three separate \textit{Task Executor} components, so that each \textit{Task Executor} corresponds to one IoT application's task. Considering the granularity level (e.g., task, service) of IoT applications in FogBus2, an application can be deployed on distributed servers for execution. The \textit{Task Executor} consists of two sub-Cs, called \textit{executor} and \textit{local logger}. The former Sub-C initiates the execution of one task and forwards the results to the next \textit{Task Executor} components if the IoT application is developed using the dependent model. It is crystal clear that in the independent model, the results will be forwarded to the \textit{Master} component for the aggregation or directly to the corresponding \textit{User} component. Besides, the \textit{local logger} sub-C records the contextual information of this task, such as its execution time.      
\end{itemize}

\subsection{Interaction Scenario}
\begin{figure}[t]
    \centering
    \includegraphics[width=\linewidth,height=8cm]{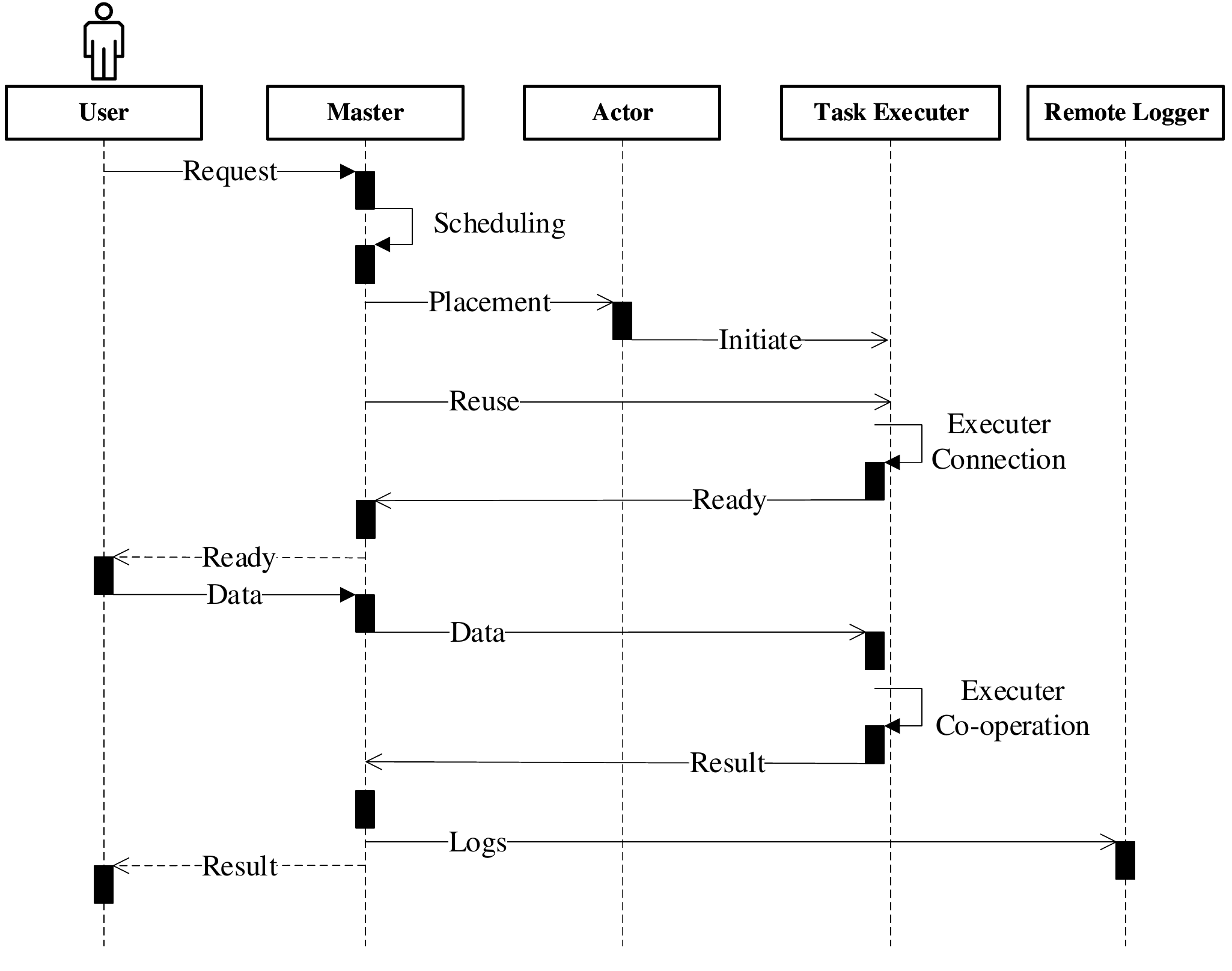}
    \caption{FogBus2 Sequence Diagram}
    \label{fig:Fogbus2SequenceDiagram}
\end{figure}
Considering the FogBus2 framework is in a ready state, Figure \ref{fig:Fogbus2SequenceDiagram} depicts the interaction of IoT users with the framework as a sequence diagram. The IoT device runs a specific \textit{User} component for each IoT application, configuring and controlling the sensing interval and aggregation of sensory data. The \textit{User} component sends a placement request to the \textit{Master} component. The \textit{Master} checks the IoT device and requested IoT application, and accordingly assigns it a unique identifier and registers it in its record. Next, the \textit{Master} calls its \textit{scheduling $\&$ scaler} sub-C to handle the current placement request. The \textit{scheduling $\&$ scaler} sub-C has the contextual information of available \textit{Actors}, \textit{Task Executor} components, IoT application, and the networking condition. Accordingly, it runs the scheduling and scaling policies to find the best possible configuration of constituent parts of an IoT application. Based on the outcome of \textit{scheduling $\&$ scaler}, two scenarios may happen. In the first scenario, if there exist no available \textit{Task Executor} components to be reused for this new request, the \textit{Master} sends the placement request to the \textit{Actor} components, selected by the scheduling mechanism. Then, the \textit{Actor} components who receive this message initiate corresponding \textit{Task Executor} components on the servers on which they are assigned. In the second scenario where there are some corresponding \textit{Task Executor} components in the cooling-off period, the \textit{Master} directly reuses those \textit{Task Executor} components, which reduces the service ready time of IoT application. When all corresponding \textit{Task Executor} components of the IoT application are ready, the \textit{Master} sends a ready message to the \textit{User} component. This message states that the service is ready, and the IoT device can start sending data. Hence, \textit{User} component sends the sensory data to the \textit{Master}, and this component forwards the sensory data to the corresponding \textit{Task Executor} components. After \textit{Task Executor} components finish their execution, the result will be forwarded to the \textit{Master}. Finally, the \textit{Master} component sends the respective logs to the \textit{Remote Logger} component, and also forwards the results to the \textit{User} component.             
\par
In addition, if the current \textit{Master} component cannot handle the placement request, the request will be forwarded to other \textit{Master} components in the environment, or a new \textit{Master} component will be initiated on a new server. The rest of the steps for handling the placement request is the same as the above-mentioned process.

\subsection{Communication Protocol}
\begin{figure}[t]
    \centering
    \includegraphics[height=8cm]{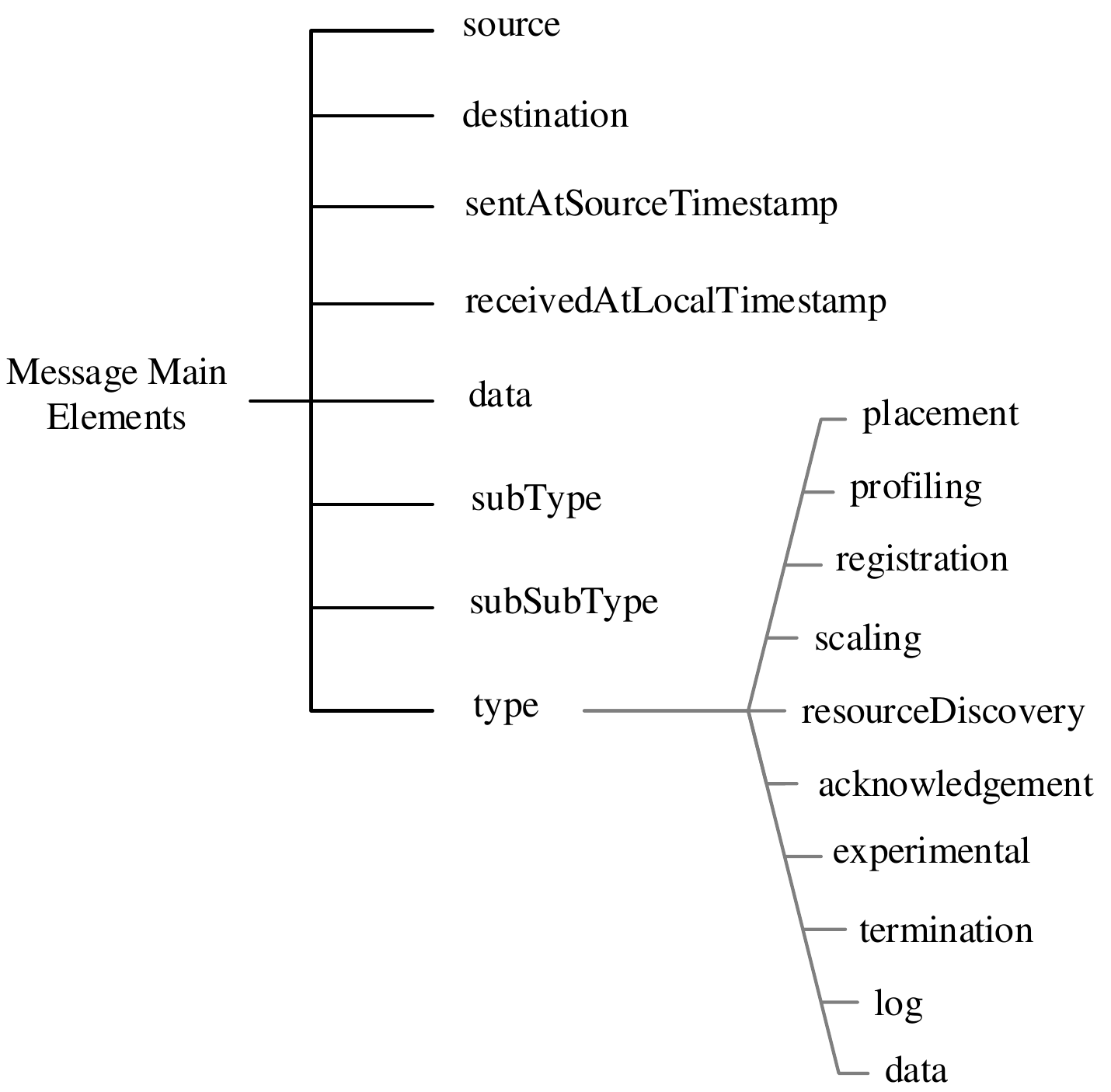}
    \caption{FogBus2 Communication Protocol Format}
    \label{fig:Fogbus2CommunicationProtocl}
\end{figure}

Different components of the FogBus2 framework can communicate together by passing messages. Therefore, understanding the communication protocol of FogBus2 is important, especially for the developers. The communication protocol of FogBus2 is implemented in JSON format and messages contain eight main elements, as depicted in Figure~\ref{fig:Fogbus2CommunicationProtocl}. 
\par
The \textit{source} and \textit{destination} are JSON objects containing the metadata of source and destination of one message, respectively. The \textit{sentAtSourceTimestamp} and \textit{receivedAtLocalTimestamp} elements are embedded to calculate the networking delay. Furthermore, each message can carry any types of information, stored in \textit{data}. Besides there are three other elements, namely \textit{type}, \textit{subType}, and \textit{subSubType}, which are used to categorize messages. There are 10 types of messages in the current version of FogBus2 framework, shown in Figure~\ref{fig:Fogbus2CommunicationProtocl}, where each \textit{type} can be further divided into 41 \textit{subType} and 5 \textit{subSubType}. Hence, \textit{type}, \textit{subType}, and \textit{subSubType} elements logically provides a hierarchical structure for the categorization of the messages. Due to the page limit we cannot describe all the messages here, however, the most important messages and their respective description are provided in Table~\ref{tab:communicationProtocolImportant}. Also, code snippet~\ref{code:messageExample} presents a sample FogBus2 message used for sending the log information ($\textit{type}=log$) of server resources ($\textit{type}=hostResources$) from an \textit{Actor} component ($\textit{source role}= \textit{Actor}$) to the \textit{Remote Logger} component ($\textit{destination role}= \textit{RemoteLogger}$). Accordingly, the message contains the resources information in the \textit{data} element.    
\begin{table}[t]
\centering
\caption{Important Communication Messages}
\label{tab:communicationProtocolImportant}
		\resizebox{1\textwidth}{!}{%
\begin{tabular}{|c|c|c|c|c|l|}
\toprule
Sender         & Receiver       & Type               & SubType              & SubSubType & \multicolumn{1}{c|}{Description}                                                                                                                                  \\ \midrule
Master         & Actor          & placement          & runTaskExecutor      & -          & \begin{tabular}[c]{@{}l@{}}Master has finished the scheduling and sends\\ this message in a no-reuse scenario\end{tabular}                                        \\ \midrule
TaskExecutor   & Master         & placement          & lookup               & -          & \begin{tabular}[c]{@{}l@{}}Task Executor requests the address of its children\\ Task Executors (in the dependent model)\end{tabular}                                \\ \midrule
Master         & TaskExecutor   & placement          & lookup               & -          & \begin{tabular}[c]{@{}l@{}}Master responds to the lookup message of\\ Task Executors\end{tabular}                                                                  \\ \midrule
TaskExecutor   & Master         & acknowledgement    & ready                & -          & \begin{tabular}[c]{@{}l@{}}Task Executor has received its children's information,\\ and use this message to acknowledge the Master\\ that it is ready\end{tabular} \\ \midrule
Master         & User           & acknowledgement    & serviceReady         & -          & \begin{tabular}[c]{@{}l@{}}When the service is ready and User can start sending\\ sensory data\end{tabular}                                                       \\ \midrule
User           & Master         & data               & sensoryData          & -          & sensoryData forwarded from the User                                                                                                                               \\ \midrule
Master         & TaskExecutor   & data               & intermediateData     & -          & \begin{tabular}[c]{@{}l@{}}Master sends sensory data to Task Executor(s) for\\ processing\end{tabular}                                                             \\ \midrule
TaskExecutor   & TaskExecutor   & data               & intermediateData     & -          & \begin{tabular}[c]{@{}l@{}}Task Executor finishes its execution and send \\ intermediate data to other Task Executor(s)\end{tabular}                                \\ \midrule
TaskExecutor   & Master         & acknowledgement    & waiting              & -          & \begin{tabular}[c]{@{}l@{}}Task Executor asks Master whether it can go \\ into the cool off period\end{tabular}                                                    \\ \midrule
Master         & TaskExecutor   & acknowledgement    & wait                 & -          & \begin{tabular}[c]{@{}l@{}}Master asks Task Executor to start its cool off\\ period immediately\end{tabular}                                                       \\ \midrule
Master         & TaskExecutor   & placement          & reuse                & -          & \begin{tabular}[c]{@{}l@{}}Master has finished the scheduling and sends\\ this message in reuse scenario\end{tabular}                                             \\ \midrule
TaskExecutor   & Master         & data               & finalResult          & -          & Task Executor sends final results to Master                                                                                                                        \\ \midrule
Master         & User           & data               & finalResult          & -          & Master sends final results to User                                                                                                                                \\ \midrule
Master A       & Master B       & scaling            & getProfiles          & -          & \begin{tabular}[c]{@{}l@{}}Master A send request to get profiles from the\\ Master B\end{tabular}                                                                 \\ \midrule
Master B       & Master A       & scaling            & profilesInfo         & -          & Master B sends profiles to Master B                                                                                                                               \\ \midrule
Master         & Actor          & scaling            & initNewMaster        & -          & Master asks Actor to initiate a new Master                                                                                                                        \\ \midrule
RemoteLogger   & Master         & log                & allResourcesProfiles & -          & \begin{tabular}[c]{@{}l@{}}This message is sent in response to requestProfiles\\ message of the Master\end{tabular}                                               \\ \midrule
Master A       & Master B       & resourcesDiscovery & requestActorsInfo    & -          & \begin{tabular}[c]{@{}l@{}}Master A asks Master B the information of  Actors\\ registered at Master B for further advertisement\end{tabular}                      \\ \midrule
Master B       & Master A       & resourcesDiscovery & actorsInfo           & -          & \begin{tabular}[c]{@{}l@{}}Master B sends its registered Actors' information\\ to Master A\end{tabular}                                                           \\ \midrule
Master         & Actor          & resourcesDiscovery & advertiseMaster      & -          & Master advertises itself to Actor                                                                                                                                 \\ \midrule
Any Components & Any Components & resourcesDiscovery & probe                & try        & \begin{tabular}[c]{@{}l@{}}Any component receiving probe message should\\ provide its component role, such as Master, Actor,\\ etc to the sender\end{tabular}     \\ \midrule
Any Components & Any Components & resourcesDiscovery & probe                & result     & \begin{tabular}[c]{@{}l@{}}The response to the probe message received from\\ one component\end{tabular}                                                           \\ \bottomrule
\end{tabular}
}
\end{table}

\renewcommand{\lstlistingname}{Code Snippet}
\begin{lstlisting}[language=Python,caption={An Example of FogBus2 Message Format},label=code:messageExample]
{'data': {'resources': {'cpu': {'cores': 8,  // Message type is log, subtype is hostResources. Thus, data contains resources 
                  'frequency': 2400.0,
                  'utilization': 0.052,
                  'utilizationPeak': 1.0},
                  'memory': {'maximum': 17179869184,
                  'utilization': 0.075,
                  'utilizationPeak': 1.0}}},
 'destination': {'addr': ['127.0.0.1', 5000],
                 'componentID': '?',
                 'hostID': 'HostID',
                 'name': 'RemoteLogger-?_127.0.0.1-5000',
                 'nameConsistent': 'RemoteLogger_HostID',
                 'nameLogPrinting': 'RemoteLogger-?_127.0.0.1-5000',
                 'role': 'RemoteLogger'},
 'receivedAtLocalTimestamp': 0.0,
 'sentAtSourceTimestamp': 1625572932123.89,
 'source': {'addr': ['127.0.0.1', 50000],
                  'componentID': '2',
                  'hostID': '127.0.0.1',
                  'name': 'Actor',
                  'nameConsistent': 'Actor_127.0.0.1',
                  'nameLogPrinting': 'Actor-2_127.0.0.1-50000_Master-?_127.0.0.1-5001',
                  'role': 'Actor'},
 'subSubType': '',
 'subType': 'hostResources',
 'type': 'log'}
\end{lstlisting}

\subsection{Main Capabilities}
In this section, we briefly describe the main capabilities of the FogBus2 framework.
\begin{itemize}
\item{\textbf{Container-enabled}:} All components of the FogBus2 framework, alongside IoT applications, are containerized. Not only does this feature enable fast deployment of IoT applications, but it also leads to faster deployment of the framework's components. Also, it brings fast portability as the containerized IoT applications and components of the framework can run smoothly on different servers. 

\item{\textbf{Multi platform support}:} In a highly heterogeneous computing environment, a wide variety of servers and IoT devices with different platforms (e.g., Intel x86, AMD, ARM) exist. To fully utilize the potential of heterogeneous servers in the cloud and/or at the edge, the containerized framework should be compatible with different platforms. To achieve this, the FogBus2 framework uses multi-arch images. Such images are built and pushed to registries with multiple variants of operating systems or CPU architectures while the image name is the same for all. Accordingly, pulling images on a server with specific architecture results in a compatible image variant for that server.

\item{\textbf{Scheduling}:} Considering available resources of heterogeneous servers and various types of IoT applications with different levels of resource requirements, the scheduling of incoming requests of IoT applications is of paramount importance. As a result, the \textit{Master} component of the FogBus2 framework is embedded with a \textit{scheduler $\&$ scaler} sub-C, which is integrated with different scheduling policies. The researchers and developers can either use the integrated policies or can develop their scheduling policies and integrate them with the \textit{scheduler $\&$ scaler} sub-C.

\item{\textbf{Dynamic scalability}:} The number of IoT devices and incoming requests varies at different times. If the number of incoming requests increases, the framework may become a bottleneck as the queuing time of incoming requests, which require scheduling and processing, increases. Hence, a dynamic scalability mechanism is embedded in the \textit{Master} component of the FogBus2 framework to dynamically scale up the \textit{Master} components as the number of incoming requests increases, which significantly reduce the queuing time of incoming requests from IoT applications. FogBus2 users can use the integrated scalability policy of the FogBus2, or develop their scalability policies.

\item{\textbf{Dynamic resource discovery}:} The highly heterogeneous and integrated computing environments, as depicted in Figure \ref{fig:heterogeneousComputingEnvironment} are considerably dynamic. This indicates that new servers may join or leave the environment due to different reasons. Furthermore, each server may run different components of the FogBus2 framework at a specific time. Hence, the FogBus2 framework offers a dynamic resource discovery mechanism to discover available servers in the environment and the containers they are running. 
This feature ensures the last-minute information of available servers and their functionalities are always accessible. 

\item{\textbf{Supporting different topology models for communication}:} IoT applications require different communication models such as client-server and peer-to-peer (P2P), just to mention a few. Accordingly, to efficiently manage the inter-component communications for different IoT applications, each containerized component of the FogBus2 framework contains a \textit{message handler} sub-C which is responsible for sending and receiving messages to/from other components. Therefore, based on the distributed message handling mechanism of the FogBus2 framework, researchers and developers can implement different communication topology models based on their IoT application scenarios.

\item{\textbf{Virtual Private Network (VPN) support}:} In the highly heterogeneous computing environment, several servers with public and private IP addresses exist. Servers with public addresses can bi-directionally communicate with each other. However, servers with private addresses cannot bi-directionally communicate with servers with public and private IP addresses. As a result, the FogBus2 put forward a P2P VPN script, working based on the Wireguard \footnote{\href{https://www.wireguard.com/}{https://www.wireguard.com/}}, as an optional feature for researchers and developers to set up a VPN among all desired servers. Among the most common VPN tools, the Wireguard has the least overhead, making it a suitable option for IoT applications, specifically real-time and latency-critical ones.       

\item{\textbf{Supporting heterogeneous IoT applications}:} FogBus2 framework supports various types of IoT applications, ranging from latency-critical and real-time IoT applications to highly computation-intensive IoT applications. Besides, it provides several ready-to-use containerized and modularized IoT applications for its users. Hence, they can simply use the current embedded IoT applications, extend current IoT applications by modifying modules of current IoT applications or integration of new modules or define their desired IoT applications from scratch.   

\item{\textbf{Distributed multi-database platform support}:} Fogbus2 framework is currently integrated with two different databases. First, it uses the containerized version of MariaDB \footnote{\href{https://mariadb.org/}{https://mariadb.org/}} which is an open-source MySQL-based database developed by the original developers of MySQL. Besides, it is integrated with Oracle autonomous database (AutoDB) \footnote{\href{https://www.oracle.com/au/autonomous-database/}{https://www.oracle.com/au/autonomous-database/}} which is an intelligent cloud-based database. AutoDB uses machine learning to automate database tuning, security, backups, updates, and other routine management tasks without human intervention. 

\item{\textbf{Reusability}:} Containerization significantly helps to reduce the deployment time of IoT applications compared to the traditional deployment techniques. However, as the number of incoming requests from different IoT devices increases, the startup time of containers, serving the IoT requests, may negatively affect the service ready time. Accordingly, the FogBus2 framework offers a configurable cooling-off period for the \textit{Task Executor} components, during which containers keep waiting for the next incoming request of the same type before stopping. This feature significantly helps to reduce the service ready time of IoT applications, specifically when the environment is crowded.         

\item{\textbf{Usability}:} FogBus2 offers a default setting for users, by which they can easily run the embedded IoT applications and test the functionality of all framework's components. Besides, users can play with several embedded options to configure IoT applications and the framework's components according to their desired scenario. In the rest of this chapter, we explain the most important options of this framework so that the users can efficiently configure the framework.  
\end{itemize}

\section{Installation of FogBus2 Framework}
FogBus2 is a new containerized framework developed by the Cloud Computing and Distributed Systems (CLOUDS) Laboratory at the University of Melbourne. As the FogBus2 framework targets both users and developers, we have provided two ways for building images of docker containers: 1) building from scratch and 2) pulling from docker hub. Accordingly, a straightforward way to install this framework is put forward.  

\subsection{Building From Scratch}
\label{setion:BuildFromScratch}
IoT developers may want to extend and configure the FogBus2 framework and define their applications on top of this framework. Hence, it is required for them to know how to build the images from scratch. In what follows, we describe this process, which is tested on Ubuntu 18.04, Ubuntu 20.04, Ubuntu 21.04, and macOS Big Sur.

\begin{enumerate}
    \item Prepare the prerequisites:
        \begin{enumerate}
        \item Install python3.9+
        \item Install pip3
        \item Install docker engine
        \item Install docker compose
        \end{enumerate}
    \item Clone/download the source code of FogBus2 framework from \href{https://github.com/Cloudslab/FogBus2}{https://github.com/Cloudslab/FogBus2} to any desired location.
    \item go to the FogBus2 folder:
    \begin{lstlisting}[language=bash,autogobble]
    $ cd fogbus2
    $ pwd
    /home/ubuntu/fogbus2
    \end{lstlisting}
    \item Install the required dependencies:
    \begin{lstlisting}[language=bash,autogobble]
    $ python3.9 -m pip install -r /containers/user/sources/requirements.txt
    \end{lstlisting}
    \item Prepare and configure the database:
    \begin{lstlisting}[language=bash,autogobble]
    $ cd containers/database/mariadb/
    $ python3.9 configure.py --create --init
    \end{lstlisting}
    \item Build all docker images:
    \begin{lstlisting}[language=bash,autogobble]
    $ pwd
    /home/ubuntu/fogbus2/demo
    $ python3.9 demo.py --buildAll
    \end{lstlisting}
\end{enumerate}

The \textit{demo.py} file automatically starts building all docker images to simplify this process. This process may take a long time to complete based on the server on which you are building the images. Besides, after any changes the developers apply to the code, the images should be rebuilt. Moreover, in distributed application scenarios, where different components should run on different servers, the components' images should be created or migrated on/to different servers. To do so, \textit{demo.py} can be configured through command-line options to only build specific images rather than creating all images. Finally, developers who are interested in extending the framework or defining new applications can use this file to understand how to create and configure their images.

\subsection{Pulling From Docker Hub}
To simply use and test the latest features of the FogBus2 framework, the multi-arch images of different components of this framework are available in the docker hub to be pulled. Although it is a faster and simpler way to run and test the FogBus2 framework and its integrated applications, users who are interested in extending and modifying this framework should build the images from scratch. In what follows, we describe the required steps to install the FogBus2 framework using uploaded images to the docker hub.
\begin{enumerate}
    \item Prepare the prerequisites:
        \begin{enumerate}
        \item Install docker engine
        \end{enumerate}
    \item Pull the docker images of Master, Actor, User, and RemoteLogger on desired servers using the following commands:
    \begin{lstlisting}[language=bash,autogobble]
    $ docker pull cloudslab/fogbus2-remote_logger && docker tag cloudslab/fogbus2-remote_logge fogbus2-remote_logger
    $ docker pull cloudslab/fogbus2-master && docker tag cloudslab/fogbus2-master fogbus2-master
    $ docker pull cloudslab/fogbus2-actor && docker tag cloudslab/fogbus2-actor fogbus2-actor
    $ docker pull cloudslab/fogbus2-user && docker tag cloudslab/fogbus2-user fogbus2-user
    \end{lstlisting}
    \item Install any desired applications by means of pulling respective docker images (i.e., \textit{Task Executor} components) of that application. To illustrate, the following command is put forward:
    \begin{enumerate}
        \item Install video-OCR application:
        \begin{lstlisting}[language=bash,autogobble]
        $ docker pull cloudslab/fogbus2-ocr && docker tag cloudslab/fogbus2-ocr fogbus2-ocr
    \end{lstlisting}    
    \end{enumerate}
\end{enumerate}

The \textit{video-OCR} application consists of one \textit{Task Executor}, called \textit{fogbus2-ocr}. However, there exist other integrated applications, which each one contains several dependent \textit{Task Executor} components. For such applications, all dependent \textit{Task Executor} components should be pulled for the proper execution of the application. 

\section{Sample FogBus2 Setup}
In this section, we describe how to configure the FogBus2 framework to run some of the currently integrated applications. We suppose that docker images are properly built or pulled on the servers, and they are ready to use.
\par
Our sample integrated computing environment consists of six CSs, tagged by \textit{A} to \textit{F}, two ESs, tagged by \textit{G} to \textit{H}, and a device playing the role of an IoT device, tagged as \textit{I}. We have used three Oracle Ampere A1 Instances ~\footnote{\href{https://www.oracle.com/au/cloud/compute/arm/}{https://www.oracle.com/au/cloud/compute/arm/}}\footnote{To reproduce this setup, you can use up to 4 Oracle Ampere A1 instances in always free Oracle Cloud Free Tier.
} and three Nectar
instances~\footnote{ARDC’s Nectar Research Cloud is an Australian federated research cloud.} to set up a multi-cloud environment. Besides, as ESs, we used Raspberrypi 4B\footnote{\href{https://www.raspberrypi.org/products/raspberry-pi-4-model-b/}{https://www.raspberrypi.org/products/raspberry-pi-4-model-b/}} and Nvidia Jetson Nano\footnote{\href{https://www.nvidia.com/en-au/autonomous-machines/embedded-systems/jetson-nano/}{https://www.nvidia.com/en-au/autonomous-machines/embedded-systems/jetson-nano/}} to set up an edge computing layer with heterogeneous resources. Our CSs have public IP addresses, while ESs do not hold public IP addresses. In this case, to integrate ESs and CSs in the FogBus2 framework VPN connection is required. Consequently, we provide a guideline and a script to simply establish a P2P VPN between all participating servers, either at the edge or at the cloud. It is crystal clear that in case all servers have public IP addresses or all components are running on one server, the VPN is not required. Table~\ref{tab:HostConfig} shows the list of servers, their computing layer, public IP addresses, private IP addresses after the establishment of the VPN connection, and the FogBus2's components running on each server. In the rest of this section, we describe how to set up a P2P VPN, assign private IP addresses to these servers, and how to run FogBus2 components on each server. As a prerequisite, make sure to open the required ports on servers.

    \begin{lstlisting}[language=bash,autogobble]
    # Required Ports for FogBus2 Components
    REMOTE_LOGGER_PORT_RANGE=5000-5000
    MASTER_PORT_RANGE=5001-5010
    ACTOR_PORT_RANGE=50000-50100
    USER_PORT_RANGE=50101-50200
    TASK_EXECUTOR_PORT_RANGE=50201-60000
    
    # Required Port for Wireguard
    WG_PORT=4999

    # Required Port for MariaDB Database
    PORT=3306
    
    \end{lstlisting}

\subsection{P2P VPN Setup}
We have used the Wireguard to set up a lightweight P2P VPN connection among all servers. In what follows, we describe how to install and configure VPN while all servers run Ubuntu as their operating system: 

\begin{table}[t]
\centering
\caption{Sample Configuration of Servers in Integrated Computing Environment}
\label{tab:HostConfig}
\resizebox{1\textwidth}{!}{%
\begin{tabular}{|c|c|c|c|c|c|c|c|}
\hline
\begin{tabular}[c]{@{}c@{}}Server\\ Tag\end{tabular} & \begin{tabular}[c]{@{}c@{}}Server\\ Name\end{tabular}     & \begin{tabular}[c]{@{}c@{}}Computing\\ Layer\end{tabular} & \begin{tabular}[c]{@{}c@{}}Public IP\\ Address\end{tabular} & \begin{tabular}[c]{@{}c@{}}Private IP\\ Address\end{tabular} & Port                                                           & \begin{tabular}[c]{@{}c@{}}Component\\ Role\end{tabular}         & \begin{tabular}[c]{@{}c@{}}Environment\\ Preparation\end{tabular}    \\ \hline
A                                                    & Oracle1                                                   & Cloud                                                     & 168.138.9.91                                                & 192.0.0.1                                                    & 5000                                                           & \begin{tabular}[c]{@{}c@{}}RemoteLogger,\\ Actor\_1\end{tabular} & \begin{tabular}[c]{@{}c@{}}docker and \\ docker-compose\end{tabular} \\ \hline
B                                                    & Oracle2                                                   & Cloud                                                     & 168.138.10.94                                               & 192.0.0.2                                                    & \begin{tabular}[c]{@{}c@{}}automatically\\ assign\end{tabular}                                                           & Actor\_2                                                           & \begin{tabular}[c]{@{}c@{}}docker and \\ docker-compose\end{tabular} \\ \hline
C                                                    & Oracle3                                                   & Cloud                                                     & 168.138.15.110                                              & 192.0.0.3                                                    & \begin{tabular}[c]{@{}c@{}}automatically\\ assign\end{tabular} & Actor\_3                                                         & \begin{tabular}[c]{@{}c@{}}docker and \\ docker-compose\end{tabular} \\ \hline
D                                                    & Nectar1                                                   & Cloud                                                     & 45.113.235.222                                              & 192.0.0.4                                                    & \begin{tabular}[c]{@{}c@{}}automatically\\ assign\end{tabular} & Actor\_4                                                         & \begin{tabular}[c]{@{}c@{}}docker and \\ docker-compose\end{tabular} \\ \hline
E                                                    & Nectar2                                                   & Cloud                                                     & 45.113.232.187                                              & 192.0.0.5                                                    & \begin{tabular}[c]{@{}c@{}}automatically\\ assign\end{tabular} & Actor\_5                                                         & \begin{tabular}[c]{@{}c@{}}docker and \\ docker-compose\end{tabular} \\ \hline
F                                                    & Nectar3                                                   & Cloud                                                     & 45.113.232.245                                              & 192.0.0.6                                                    & \begin{tabular}[c]{@{}c@{}}automatically\\ assign\end{tabular} & Actor\_6                                                         & \begin{tabular}[c]{@{}c@{}}docker and \\ docker-compose\end{tabular} \\ \hline
G                                                    & \begin{tabular}[c]{@{}c@{}}RPi 4B\\ 4GB\end{tabular}      & Edge                                                      & -                                                           & 192.0.0.7                                                    & \begin{tabular}[c]{@{}c@{}}automatically\\ assign\end{tabular} & Actor\_7                                                         & \begin{tabular}[c]{@{}c@{}}docker and \\ docker-compose\end{tabular} \\ \hline
H                                                    & \begin{tabular}[c]{@{}c@{}}Jetson Nano\\ 4GB\end{tabular} & Edge                                                      & -                                                           & 192.0.0.8                                                    & 5001 & Master                                                         & \begin{tabular}[c]{@{}c@{}}docker and \\ docker-compose\end{tabular} \\ \hline
I                                                    & \begin{tabular}[c]{@{}c@{}}VM on a \\ Laptop\end{tabular} & IoT                                                       & -                                                           & 192.0.0.9                                                    & \begin{tabular}[c]{@{}c@{}}automatically\\ assign\end{tabular} & User                                                             & Python3.9                                                            \\ \hline
\end{tabular}
}
\end{table}

\begin{enumerate}
    \item Install Wireguard on all servers:
    \begin{lstlisting}[language=bash,autogobble]
    $ sudo apt update
    $ sudo apt install wireguard
    $ wg --version
    wireguard-tools v1.0.20200513 - https://git.zx2c4.com/wireguard-tools/
    \end{lstlisting}
    \item Simply configure the Wireguard on servers using our auto-generating script:
    \begin{enumerate}
        \item Specify server information on \textit{hostIP.csv}:
        \begin{lstlisting}[language=bash,autogobble]
        $ pwd
        /home/ubuntu/fogbus2
        $ cd config/host/
        $ cat config/host/hostIP.csv
        hostname, publicIP
        oracle1, 168.138.9.91
        oracle2, 168.138.10.94
        oracle3, 168.138.15.110
        nectar1, 45.113.235.222
        nectar2, 45.113.232.187
        nectar3, 45.113.232.245
        rpi-4B-2G,
        JetsonNano-4G,
        VM-laptop,
        \end{lstlisting}  
        \item Automatically generate Wireguard configuration files:
        \begin{lstlisting}[language=bash,autogobble]
        $ pwd
        /home/ubuntu/fogbus2
        $ cd scripts/wireguard/
        $ python3.9 generateConf.py
        ...
        ==========================================
        hostname WireguardIP
        oracle1 192.0.0.1
        oracle2 192.0.0.2
        oracle3 192.0.0.3
        nectar1 192.0.0.4
        nectar2 192.0.0.5
        nectar3 192.0.0.6
        rpi-4B-2G-4B 192.0.0.7
        JetsonNano-4G 192.0.0.8
        VM-laptop 192.0.0.9
        ==========================================
        [*] Generated Wireguard config for oracle1: /path/to/proj/output/wireguardConfg/oracle1/wg0.conf 
        ...
        ==========================================
        \end{lstlisting}
        \item Copy obtained configuration files to \textit{/etc/wireguard/wg0.conf} of each server, respectively.
        \item Run Wireguard on each server:
        \begin{lstlisting}[language=bash,autogobble]
        $ sudo wg-quick up /etc/wireguard/wg0.conf && sudo wg
        \end{lstlisting}
    \end{enumerate}
    \item Test the P2P VPN connection using \textit{Ping} command and private IP addresses.
    \item If \textit{Ping} command does not properly work, make sure to open the configured Wireguard port on all servers. In FogBus2, the default port of Wireguard is set to UDP 4999, which can be changed from \textit{/home/ubuntu/fogbus2/config/network.env}.  
    \begin{lstlisting}[language=bash,autogobble]
    #Install, enable and configure Firewalld
    $ sudo apt update
    $ sudo apt install firewalld
    $ sudo systemctl enable firewalld
    $ sudo firewall-cmd --state
    $ sudo firewall-cmd --permanent --zone=public  --add-port=22/tcp --add-port=53/tcp --add-port=3306/tcp --add-port=4999/udp --add-port=5000-5010/tcp --add-port=5000-60000/tcp
    $ sudo firewall-cmd --reload    
    \end{lstlisting}
\end{enumerate}

\subsection{Running FogBus2 Components}
Considering Table~\ref{tab:HostConfig}, the FogBus2 components should run on different servers. Also, each server may run several components simultaneously and play different roles, similar to the server \textit{A}. In what follows, we describe how to run these components and provide respective commands.

\begin{enumerate}
    \item Starting \textit{RemoteLogger} component:
    \begin{enumerate}
        \item Configure database credentials in \textit{containers/remoteLogger/sources/.mysql.env}
        \begin{lstlisting}[language=bash,autogobble]
        $ pwd
        /home/ubuntu/fogbus2
        $ cat containers/remoteLogger/sources/.mysql.env
        HOST=192.0.0.1
        PORT=3306
        USER=root
        PASSWORD=passwordForRoot
        \end{lstlisting}
        \item Run \textit{RemoteLogger} component on server \textit{A}
        \begin{lstlisting}[language=bash,autogobble]
        $ pwd
        /home/ubuntu/fogbus2
        $ cd containers/remoteLogger
        $ docker-compose run --rm --name TempContainerName fogbus2-remote_logger --bindIP 192.0.0.1 --containerName TempContainerName
        \end{lstlisting}
    \end{enumerate}
    
    \item Starting \textit{Master} component on server \textit{H}. The \textit{schedulerName} specifies the name of scheduling policy used by this \textit{Master} component. Hence, in computing environments with multiple \textit{Master} components, each \textit{Master} component can be separately configured to run different scheduling policies:
    \begin{enumerate}
        \item On server \textit{H}, configure database credentials:
            \begin{lstlisting}[language=bash,autogobble]
            $ pwd
            /home/ubuntu/fogbus2/containers/master/sources/
            cat .mysql.env
            HOST=192.0.0.1
            PORT=3306
            USER=root
            PASSWORD=passwordForRoot
            \end{lstlisting}
        \item Run \textit{Master} component on the server \textit{H}:
        \begin{lstlisting}[language=bash,autogobble]
        $ pwd
        /home/ubuntu/fogbus2
        $ cd containers/master
        $ docker-compose run --rm --name TempContainerName fogbus2-master --bindIP 192.0.0.8 --bindPort 5001 --remoteLoggerIP 192.0.0.1 --remoteLoggerPort 5000 --schedulerName OHNSGA --containerName TempContainerName
        \end{lstlisting}
    \end{enumerate}
    
    \item Starting \textit{Actor} components:
    \begin{enumerate}
        \item Run \textit{Actor} component on server \textit{A} 
    
    \begin{lstlisting}[language=bash,autogobble]
        $ pwd
        /home/ubuntu/fogbus2
        $ cd containers/actor
        $ docker-compose run --rm --name TempContainerName fogbus2-actor --bindIP 192.0.0.1  --remoteLoggerIP 192.0.0.1 --remoteLoggerPort 5000 --masterIP 192.0.0.8 --masterPort 5001 --containerName TempContainerName
    \end{lstlisting}
    \item Run \textit{Actor} components on servers \textit{B} to \textit{G} using the above-mentioned command. You need to modify the value of \textit{bindIP} option for each server and use the respective private IP address of that server. For instance, to run the \textit{Actor} component on server \textit{B}:
    \begin{lstlisting}[language=bash,autogobble]
        $ pwd
        /home/ubuntu/fogbus2
        $ cd containers/actor
        $ docker-compose run --rm --name TempContainerName fogbus2-actor --bindIP 192.0.0.1  --remoteLoggerIP 192.0.0.2 --remoteLoggerPort 5000 --masterIP 192.0.0.8 --masterPort 5001 --containerName TempContainerName
    \end{lstlisting}
    \end{enumerate}
    
    \item Considering the IoT application, the \textit{User} component can run with different \textit{applicationName} options. Current version of FogBus2 comes with two integrated IoT applications, called \textit{VideoOCR}~\cite{deng2021fogbus2} and \textit{GameOfLifeParallelized}~\cite{deng2021fogbus2}, while in the rest of this book chapter we design and implement more IoT applications to describe how to define new IoT applications.
    \begin{enumerate}
        \item Run \textit{User} component for \textit{VideoOCR} application on server \textit{I}. The \textit{videoPath} shows the the address of input video to feed into VideoOCR algorithm.
        \begin{lstlisting}[language=bash,autogobble]
        $ pwd
        /home/ubuntu/fogbus2
        $ cd containers/user/sources
        $ python3.9 user.py --bindIP 192.0.0.9 --masterIP 192.0.0.8 --masterPort 5001 --remoteLoggerIP 192.0.0.1 --remoteLoggerPort 5000 --applicationName VideoOCR --applicationLabel 720 --videoPath /path/to/video.mp4
        \end{lstlisting}
        \item Run \textit{User} component for \textit{GameOfLifeParallelized} application on server \textit{I}.
        \begin{lstlisting}[language=bash,autogobble]
        $ pwd
        /home/ubuntu/fogbus2
        $ cd containers/user/sources
        $ python user.py --bindIP 192.0.0.9 --masterIP 192.0.0.8 --masterPort 5001 --remoteLoggerIP 192.0.0.1 --remoteLoggerPort 5000 --applicationName GameOfLifeParallelized --applicationLabel 480
        \end{lstlisting}
    \end{enumerate}
\end{enumerate}

\section{Extending FogBus2 Framework and New IoT Applications}
In this section, we describe how to implement and integrate new IoT applications in the FogBus2 framework. Secondly, we put forward a new scheduling algorithm and demonstrate its integration procedure with this framework.

\subsection{Implementation of New IoT Applications}
Every containerized IoT application can be implemented and integrated with the FogBus2 framework. Alongside the implementation of a new IoT application, there are several required steps to follow in order to implement and integrate the new IoT application with the FogBus2 framework, such as building docker images and defining dependencies between different tasks. In what follows, we describe a straightforward mathematical application, how to implement it, and how to integrate it with the FogBus2 framework.
\par
\begin{figure}[!t]
    \centering
    \includegraphics[height=7cm]{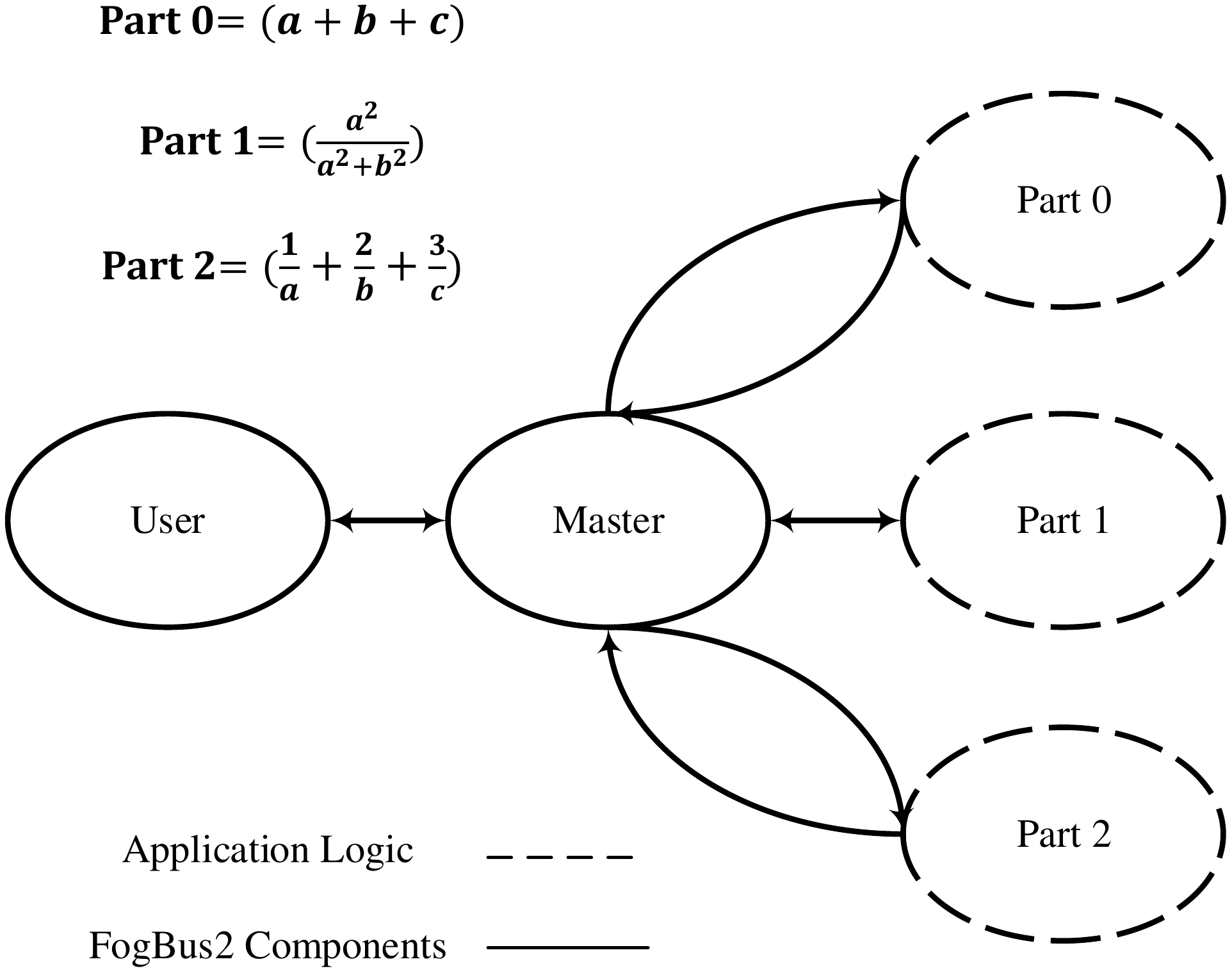}
    \caption{A Logical Model of a New Application}
    \label{fig:NewApplication}
\end{figure}
Figure~\ref{fig:NewApplication} shows a new application to be implemented with the FogBus2 framework. This mathematical application contains three different tasks, called \textit{Part 0}, \textit{Part 1}, and \textit{Part 2}, that can be executed in parallel, and it requires three inputs as \textit{a, b, c}. To integrate this application into the FogBus2 framework, three tasks should be dockerized and prepared to be integrated as \textit{Task Executor} components. Besides, we need a \textit{User} component to receive inputs (using \textit{Sensor} sub-C) and show outputs (using \textit{Actuator} sub-C). The input will be forwarded to the \textit{Master} component of the framework, and this component forwards inputs to corresponding \textit{Task Executor} components based on the outcome of the scheduling algorithm. The following steps demonstrate how to implement and integrate the new application with the FogBus2 framework:
\begin{enumerate}
    \item Create three python files as three different tasks with the desired naming convention. We name these files as \textit{naiveFormula0.py}, \textit{naiveFormula1.py}, and \textit{naiveFormula2.py} which contain the logic of tasks \textit{Part 0}, \textit{Part 1}, and \textit{Part 2}, respectively.    
    \begin{lstlisting}[language=bash,autogobble]
        $ pwd
        /home/ubuntu/fogbus2
        $ cd containers/taskExecutor/sources/utils/taskExecutor/tasks
        $ > naiveFormula0.py
        $ > naiveFormula1.py
        $ > naiveFormula2.py
    \end{lstlisting}
    \item Edit the corresponding python files of each task and insert the required logic. For each task, a unique identifier \textit{taskID} is required.
        \begin{enumerate}
        \item The logic of task \textit{naiveFormula0.py}: 
        
        \begin{lstlisting}[language=python,autogobble]
        $ nano naiveFormula0.py
            from .base import BaseTask
            
            class NaiveFormula0(BaseTask):
                def __init__(self):
                    super().__init__(taskID=108, taskName='NaiveFormula0')
            
                def exec(self, inputData):
                    a = inputData['a']
                    b = inputData['b']
                    c = inputData['c']
            
                    result = a + b + c
                    inputData['resultPart0'] = result
                    
                    return inputData
        \end{lstlisting}
        \item The logic of task \textit{naiveFormula1.py}: 
        
        \begin{lstlisting}[language=python,autogobble]
        $ nano naiveFormula1.py
            from .base import BaseTask

            class NaiveFormula1(BaseTask):
                def __init__(self):
                    super().__init__(taskID=109, taskName='NaiveFormula1')
            
                def exec(self, inputData):
                    a = inputData['a']
                    b = inputData['b']
                    c = inputData['c']
            
                    result = a * a / (b * b + c * c)
                    inputData['resultPart1'] = result
            
                    return inputData
        \end{lstlisting}
                \item The logic of task \textit{naiveFormula2.py}: 
        
        \begin{lstlisting}[language=python,autogobble]
        $ nano naiveFormula2.py
            from .base import BaseTask
            
            class NaiveFormula2(BaseTask):
                def __init__(self):
                    super().__init__(taskID=110, taskName='NaiveFormula2')
            
                def exec(self, inputData):
                    a = inputData['a']
                    b = inputData['b']
                    c = inputData['c']
            
                    result = 1 / a + 2 / b + 3 / c
                    inputData['resultPart2'] = result
                    return inputData
        \end{lstlisting}
        \item The return value of \textit{exec} functions in the above mentioned tasks will be managed by \textit{Task Executor}. If it is none, the return value will be ignored, otherwise, it will be forwarded to next \textit{Task Executor} components based on the specified dependencies among tasks.
        \end{enumerate}
        
        \item Configure arguments:
        \begin{enumerate}
        \item Configure \textit{\_\_init\_\_.py}:
        \begin{lstlisting}[language=python,autogobble]
        $ pwd 
        /home/ubuntu/fogbus2/containers/taskExecutor/sources/utils/taskExecutor/tasks
        $ nano containers/taskExecutor/sources/utils/taskExecutor/tasks/__init__.py
        
        from .base import BaseTask
        ...
        from .naiveFormula0 import NaiveFormula0
        from .naiveFormula1 import NaiveFormula1
        from .naiveFormula2 import NaiveFormula2
        ...
        \end{lstlisting}
        \item Configure \textit{initTask.py}:
        \begin{lstlisting}[language=python,autogobble]
        $ pwd 
        /home/ubuntu/fogbus2/containers/taskExecutor/sources/utils/taskExecutor/tools/initTask.py
        $ nano containers/taskExecutor/sources/utils/taskExecutor/tasks/__init__.py
        
        from typing import Union
        from ..tasks import *
        
        def initTask(taskName: str) -> Union[BaseTask, None]:
            task = None
            if taskName == 'FaceDetection':
                task = FaceDetection()
            ...
            elif taskName == 'NaiveFormula0':
                task = NaiveFormula0()
            elif taskName == 'NaiveFormula1':
                task = NaiveFormula1()
            elif taskName == 'NaiveFormula2':
                task = NaiveFormula2()
        
            return task
        \end{lstlisting}
        \end{enumerate}
        \item Prepare docker images:
        \begin{enumerate}
            
            \item Prepare the required libraries:
            \begin{lstlisting}[language=bash,autogobble]
            $ pwd
            /home/ubuntu/fogbus2/containers/taskExecutor/sources
            $ cat requirements.txt
            
            psutil
            docker
            python-dotenv
            pytesseract
            editdistance
            six
            \end{lstlisting}            
            \item Create dockerfiles: For each of the tasks, a docker file should be created. Considering \textit{NaiveFormula0}: 
            \begin{lstlisting}[language=python,autogobble]
            $ pwd 
            /home/ubuntu/fogbus2/containers/taskExecutor/dockerFiles/NaiveFormula0
            
            $ nano Dockerfile
            
            # Base
            FROM python:3.9-alpine3.14 as base
            FROM base as builder
            
            ## Dependencies
            RUN apk update
            RUN apk add --no-cache \
                build-base clang clang-dev ninja cmake ffmpeg-dev \
                freetype-dev g++ jpeg-dev lcms2-dev libffi-dev \
                libgcc libxml2-dev libxslt-dev linux-headers \
                make musl musl-dev openjpeg-dev openssl-dev \
                zlib-dev curl freetype gcc6 jpeg libjpeg \
                openjpeg tesseract-ocr zlib unzip openjpeg-tools
            
            RUN python -m pip install --retries 100 --default-timeout=600  --no-cache-dir --upgrade pip
            RUN python -m pip install --retries 100 --default-timeout=600  numpy --no-cache-dir
            
            ## OpenCV Source Code
            WORKDIR /workplace
            RUN cd /workplace/ \
                && curl -L "https://github.com/opencv/opencv/archive/4.5.1.zip" -o opencv.zip \
                && curl -L "https://github.com/opencv/opencv_contrib/archive/4.5.1.zip" -o opencv_contrib.zip \
                && unzip opencv.zip \
                && unzip opencv_contrib.zip \
                && rm opencv.zip opencv_contrib.zip
            
            ## Configure
            RUN cd /workplace/opencv-4.5.1 \
                && mkdir -p build && cd build \
                && cmake \
                    -DOPENCV_EXTRA_MODULES_PATH=../../opencv_contrib-4.5.1/modules \
                    -DBUILD_NEW_PYTHON_SUPPORT=ON \
                    -DBUILD_opencv_python3=ON \
                    -DHAVE_opencv_python3=ON \
                    -DPYTHON_DEFAULT_EXECUTABLE=$(which python) \
                    -DBUILD_TESTS=OFF \
                    -DWITH_FFMPEG=ON \
                    ../
            
            ## Compile
            
            RUN cd /workplace/opencv-4.5.1/build && make -j $(nproc)
            RUN cd /workplace/opencv-4.5.1/build && make install
            
            ## Python libraries
            COPY ./sources/requirements.txt /install/requirements.txt
            RUN python -m pip install --retries 100 --default-timeout=600  \
                --prefix=/install \
                --no-cache-dir \
                -r /install/requirements.txt
            
            ## Copy files
            FROM base
            COPY --from=builder /install /usr/local
            COPY ./sources/ /workplace
            
            ## Install OpenCV
            COPY  --from=builder /usr/local/ /usr/local/
            COPY --from=builder /usr/lib/ /usr/lib/
            
            # Hostname
            RUN echo "NaiveFormula0" > /etc/hostname
            
            # Run NaiveFormula0
            WORKDIR /workplace
            ENTRYPOINT ["python", "taskExecutor.py"]          
            \end{lstlisting}
        \item Create docker files for \textit{NaiveFormula1} and \textit{NaiveFormula2} similar to \textit{NaiveFormula0}, as described in step (b). 
        \item Create docker-compose files: For each of the tasks, a docker-compose file should be created. Consideting \textit{NaiveFormula0}:
        \begin{lstlisting}[language=bash,autogobble]
            $ pwd
            /home/ubuntu/fogbus2/containers/taskExecutor/dockerFiles/NaiveFormula0
            $ nano docker-compose.yml
            
            version: '3'

            services:
            
              fogbus2-naive_formula0:
                image: fogbus2-naive_formula0
                build:
                  context: ../../
                  dockerfile: dockerFiles/NaiveFormula0/Dockerfile
                environment:
                  PUID: 1000
                  PGID: 1000
                  TZ: Australia/Melbourne
                network_mode:
                  host
        \end{lstlisting}
        
        \item Create docker-compose files for \textit{NaiveFormula1} and \textit{NaiveFormula2} similar to \textit{NaiveFormula0}, as described in step (d).
        \item Build docker images: The docker images corresponding to the tasks of new application can be built using the provided automated script (\textit{demo.py}), similar to step (6) in section~\ref{setion:BuildFromScratch}.
        \begin{lstlisting}[language=bash,autogobble]
            $ pwd
            /home/ubuntu/fogbus2/demo
            $ python3.9 demo.py --buildAll
        \end{lstlisting}
        \item Verify new docker images:
            
        \end{enumerate}
        \begin{lstlisting}[language=bash,autogobble]
        $ docker images

        REPOSITORY                TAG       IMAGE ID       CREATED              SIZE
        ...
        fogbus2-naive_formula1    latest    5e9ad6999801   2 minutes ago        xxx
        fogbus2-naive_formula0    latest    74cfbb128699   2 minutes ago        xxx
        fogbus2-naive_formula2    latest    924d6bc0f281   3 minutes ago        xxx
        ...
        \end{lstlisting}
    \item Prepare \textit{User} side code:
        \begin{lstlisting}[language=python,autogobble]
        $ pwd
        /home/ubuntu/fogbus2/containers/user/sources/utils/user/applications
        $ nano naiveFormulaParallelized.py
        
        from time import time
        from pprint import pformat
        from .base import ApplicationUserSide
        from ...component.basic import BasicComponent
        
        
        class NaiveFormulaParallelized(ApplicationUserSide):
        
            def __init__(
                    self,
                    videoPath: str,
                    targetHeight: int,
                    showWindow: bool,
                    basicComponent: BasicComponent):
                super().__init__(
                    appName='NaiveFormulaParallelized',
                    videoPath=videoPath,
                    targetHeight=targetHeight,
                    showWindow=showWindow,
                    basicComponent=basicComponent)
        
            def prepare(self):
                pass
        
            def _run(self):
                self.basicComponent.debugLogger.info(
                    'Application is running: %s', self.appName)
        
                # get user input of a, b, and c
                print('a = ', end='')
                a = int(input())
                print('b = ', end='')
                b = int(input())
                print('c = ', end='')
                c = int(input())
        
                inputData = {
                    'a': a,
                    'b': b,
                    'c': c
                }
        
                # put it in to data uploading queue
                self.dataToSubmit.put(inputData)
                lastDataSentTime = time()
                self.basicComponent.debugLogger.info(
                    'Data has sent (a, b, c): %.2f, %.2f, %.2f', a, b, c)
        
                # wait for all the 4 results
                while True:
                    result = self.resultForActuator.get()
        
                    responseTime = (time() - lastDataSentTime) * 1000
                    self.responseTime.update(responseTime)
                    self.responseTimeCount += 1
        
                    if 'finalResult' in result:
                        break
        
                for key, value in result.items():
                    result[key] = '%.4f' % value
                self.basicComponent.debugLogger.info(
                    'Received all the 4 results: \r\n%s', pformat(result))
        \end{lstlisting}
        \item Define dependencies among tasks of a new application in the database. Considering MariaDB is running on 192.0.0.1 as an example:
        
        \begin{enumerate}
            \item Connect to the database:
            \begin{lstlisting}[language=bash,autogobble]
            $ mysql -h 192.0.0.1 -uroot -p
            Enter password:
            \end{lstlisting}
            \item The \textit{EntryTasks} contains the root tasks of this application, where the sensory data should be forwarded.
            \begin{lstlisting}[language=bash,autogobble]
            mysql> SELECT entryTasks FROM FogBus2_Applications.applications WHERE name='NaiveFormulaParallelized';
            
            [
                "NaiveFormula0",
                "NaiveFormula1",
                "NaiveFormula2"
            ]
            \end{lstlisting}
            \item The \textit{TaskWithDependency} contains the dependencies among tasks. For each task, we define an array of \textit{parents} and \textit{children}, representing predecessor and successor tasks.
            \begin{lstlisting}[language=bash,autogobble]
            mysql> SELECT tasksWithDependency FROM FogBus2_Applications.applications WHERE name='NaiveFormulaParallelized';
            
            {
            "NaiveFormula0": {
                "parents": [
                    "Sensor"
                ],
                "children": [
                    "Actuator"
                ]
            },
            "NaiveFormula1": {
                "parents": [
                    "Sensor"
                ],
                "children": [
                    "Actuator"
                ]
            },
            "NaiveFormula2": {
                "parents": [
                    "Sensor"
                ],
                "children": [
                    "Actuator"
                ]
            }
            }
            \end{lstlisting}
            \item Considering the FogBus2 framework is running, the \textit{NaiveFormulaParallelized} can be executed using the following command:
            \begin{lstlisting}[language=bash,autogobble]
            $ pwd
            /home/ubuntu/fogbus2/containers/user/sources
            
            $ python user.py --bindIP 192.0.0.9 --masterIP 192.0.0.2 --masterPort 5001 --remoteLoggerIP 192.0.0.1 --remoteLoggerPort 5000 --applicationName NaiveFormulaParallelized
            \end{lstlisting}
        \end{enumerate}
    \end{enumerate}
    
\par
Table~\ref{tab:listOfApplications} presents the list of all applications that have been currently implemented and integrated with FogBus2 framework. The \textit{VideoOCR} and \textit{GameOfLifePyramid} applications were implemented in the main paper, while the \textit{FaceDetection}, \textit{ColorTracking}, \textit{GameOfLifeSerialized}, \textit{GameOfLifeParallelized}, \textit{NaiveFormulaSerialized}, and \textit{NaiveFormulaParallelized} are implemented and integrated as an extension in this book chapter. Due to page-limit, we only described one of these applications (i.e., \textit{NaiveFormulaParallelized}). The required steps for defining and integration of all applications are similar to what is described in this section, while the logic of each application is different.

\begin{table}[!t]
\centering
\caption{The List of All Implemented and Integrated Applications with FogBus2}
\label{tab:listOfApplications}
\resizebox{1\textwidth}{!}{%
\begin{tabular}{|c|c|c|}
\hline
\textbf{Application Name} & \textbf{Description}                                                                                                                                                         & \textbf{\begin{tabular}[c]{@{}c@{}}Application Logic\\ Tasks\end{tabular}}                                       \\ \hline
FaceDetection             & \begin{tabular}[c]{@{}c@{}}Detecting human face from video stream, either realtime or from\\ recorded files\end{tabular}                                                     & face\_detection                                                                                                  \\ \hline
ColorTracking             & \begin{tabular}[c]{@{}c@{}}Tracking colors from video stream, either realtime or from recorded\\ files. The target color can be dynamically configured via GUI.\end{tabular} & color\_tracking                                                                                                  \\ \hline
VideoOCR                  & \begin{tabular}[c]{@{}c@{}}Recognizing text from a video file. It automatically picks up\\ key frames.\end{tabular}                                                          & \begin{tabular}[c]{@{}c@{}}blur\_and\_p\_hash,\\ ocr\end{tabular}                                                \\ \hline
GameOfLifeSerialized      & \begin{tabular}[c]{@{}c@{}}Conway's Game of Life. The tasks process grids \\ (with different sizes) one by one.\end{tabular}                                                 & \begin{tabular}[c]{@{}c@{}}GameOfLife0 to\\ GameOfLife62\end{tabular}                                            \\ \hline
GameOfLifeParallelized    & \begin{tabular}[c]{@{}c@{}}Conway's Game of Life. The tasks process grids\\ (with different sizes) in parallel.\end{tabular}                                                 & \begin{tabular}[c]{@{}c@{}}GameOfLife0 to\\ GameOfLife62\end{tabular}                                            \\ \hline
GameOfLifePyramid         & \begin{tabular}[c]{@{}c@{}}Conway's Game of Life. The tasks process grids\\ (with different sizes) in a pyramid dependency.\end{tabular}                                     & \begin{tabular}[c]{@{}c@{}}GameOfLife0 to\\ GameOfLife62\end{tabular}                                            \\ \hline
NaiveFormulaSerialized    & A Naive Formula. Tasks process different parts of formula one by one.                                                                                                               & \begin{tabular}[c]{@{}c@{}}naive\_formula0,\\ naive\_formula1,\\ naive\_formula2,\\ naive\_formula3\end{tabular} \\ \hline
NaiveFormulaParallelized  & A Naive Formula. Tasks process different parts of formula in parallel.                                                                                                        & \begin{tabular}[c]{@{}c@{}}naive\_formula0,\\ naive\_formula1,\\ naive\_formula2\end{tabular}                    \\ \hline
\end{tabular}
}
\end{table}

\subsection{Implementation of New Scheduling Policy}
One of the most important challenges for resource management in edge and cloud data centers is the proper scheduling of incoming application requests. FogBus2 provides a straightforward mechanism for the scheduling of various types of IoT applications. Different scheduling policies can be implemented and integrated with the FogBus2 framework with different scheduling goals, such as optimizing application response time, energy consumption, the monetary cost of resources, or a combination of any of these goals, just to mention a few. As a guideline, we put forward a new scheduling policy and describe how to integrate it with the FogBus2 framework.
\par
To simplify the process of new policy integration, a \textit{BaseScheduler} class is provided in \textit{containers/master/sources/utils/master/scheduler/base.py}. Users should inherit from \textit{BaseScheduler} class and override the \textit{$\_schedule$} method based on their desired goals. Besides, if the utilization of the current \textit{Master} component, which is responsible for the scheduling of IoT applications, goes beyond a threshold, the new application request should be forwarded to another \textit{Master} component. The \textit{getBestMaster} method handles this process and can be overridden with different policies for the selection of another \textit{Master}. Finally, users, who are interested in augmenting scaling features to their technique, can implement a scaling policy using the \textit{prepareScaler} method. The following steps describe how to define and integrate a new scheduling policy:

\begin{enumerate}
    \item Navigate to containers/master/sources/utils/master/scheduler/policies, and create a new file named schedulerRankingBased.py:
    \begin{lstlisting}[language=bash,autogobble]
        $ pwd
        /home/ubuntu/fogbus2/containers/master/sources/utils/master/scheduler/policies
        $ > schedulerRankingBased.py
    \end{lstlisting}
    \item Implement the policy in the \textit{schedulerRankingBased.py}. The \textit{$\_schedule$} contains the logic of scheduling policy.      
    \begin{lstlisting}[language=python,autogobble]
        $ cat schedulerRankingBased.py
        
        from random import randint
        from time import time
        from typing import List
        from typing import Union
        
        from ..base import BaseScheduler as SchedulerPolicy
        from ..baseScaler.base import Scaler
        from ..baseScaler.policies.scalerRandomPolicy import ScalerRandomPolicy
        from ..types import Decision
        from ...registry.roles.actor import Actor
        from ...registry.roles.user import User
        from ....types import Component
        
        
        class SchedulerRankingBased(SchedulerPolicy):
            def __init__(
                    self,
                    isContainerMode: bool,
                    *args,
                    **kwargs):
                """
                :param isContainerMode: Whether this component is running in container
                :param args:
                :param kwargs:
                """
                super().__init__('RankingBased', isContainerMode, *args, **kwargs)
        
            def _schedule(self, *args, **kwargs) -> Decision:
                """
                :param args:
                :param kwargs:
                :return: A decision object
                """
                user: User = kwargs['user']
                allActors: List[Actor] = kwargs['allActors']
                # Get what tasks are required
                taskNameList = user.application.taskNameList
        
                startTime = time()
                indexSequence = ['' for _ in range(len(taskNameList))]
                indexToHostID = {}
                
                # Ranking of tasks belonging to an application 
                rankedTasksList = self.rankApplicationTasks(
                indexSequence, **kwargs) 
                indexToHostID = self.tasksAssignment(
                rankedTasksList, allActors, **kwargs)
    
                schedulingTime = (time() - startTime) * 1000
        
                # Create a decision object
                decision = Decision(
                    user=user,
                    indexSequence=rankedTasksList,
                    indexToHostID=indexToHostID,
                    schedulingTime=schedulingTime
                )
                # A simple example of cost estimation  
                decision.cost = self.estimateCost(decision, **kwargs)
                return decision
        
            @staticmethod
            def estimateCost(decision: Decision, **kwargs) -> float:
                # You may develop your own with the following used values
                from ..estimator.estimator import Estimator
                # Get necessary params from the key args
                user = kwargs['user']
                master = kwargs['master']
                systemPerformance = kwargs['systemPerformance']
                allActors = kwargs['allActors']
                isContainerMode = kwargs['isContainerMode']
                # Init the estimator
                estimator = Estimator(
                    user=user,
                    master=master,
                    systemPerformance=systemPerformance,
                    allActors=allActors,
                    isContainerMode=isContainerMode)
                indexSequence = [int(i) for i in decision.indexSequence]
                # Estimate the cost
                estimatedCost = estimator.estimateCost(indexSequence)
                return estimatedCost
        
            def getBestMaster(self, *args, **kwargs) -> Union[Component, None]:
                """
        
                :param args:
                :param kwargs:
                :return: A Master used to ask the user to request when this Master is busy
                """
                user: User = kwargs['user']
                knownMasters: List[Component] = kwargs['knownMasters']
                mastersNum = len(knownMasters)
                if mastersNum == 0:
                    return None
                return knownMasters[randint(0, mastersNum - 1)]
        
            def prepareScaler(self, *args, **kwargs) -> Scaler:
                # Create a scaler object and return
                scaler = ScalerRandomPolicy(*args, **kwargs)
                return scaler
            
    \end{lstlisting}
    First, we retrieve the information of \textit{user} and all available actors \textit{allActors} (lines 36-37). Then, the tasks corresponding to the requested application are retrieved and stored in \textit{taskNameList} (line 39). The \textit{rankApplicationTasks} considers the dependency model of tasks (if any) and satisfies the dependency among tasks while defining an order for the tasks that can be executed in parallel. Several ranking policies can be defined for this method, however, in this version, we consider the average execution time of tasks on different servers as criteria for ranking. Hence, among tasks that can be executed in parallel, the tasks with higher execution time receive higher priority. This eventually helps to reduce the overall response time of the application (lines 46-47). Next, \textit{tasksAssignment} method receives the ordered \textit{rankedTasksList} and assigns a proper actor to each task to minimize its execution time (line 48-49). According to the scheduling decision, a \textit{decision} object will be created, storing the ordered list of the application's tasks, the list of server/host mapping, scheduling time, and the cost of scheduling, to be returned (lines 54-59). To illustrate how the execution cost of each task and overall response time of one application can be estimated, a \textit{estimateCost} method is defined (lines 65-84). As mentioned above, the \textit{getBestMaster} and \textit{prepareScaler} also can be defined in \textit{schedulerRankingBased.py}. To reduce the complexity, these methods are working based on random policy.       
    
    \item The new scheduling policy can be added to the \textit{schedulerName} options by manipulating the \textit{initSchedulerByName} method of the \textit{containers/master/sources/utils/master/scheduler/tools/initSchedulerByName.py}. There exist names of several scheduling policies currently integrated with the FogBus2 framework. The name of new scheduling policies can be added after the names of existing scheduling policies.
            \begin{lstlisting}[language=python,autogobble]
            $ pwd
            /home/ubuntu/fogbus2/containers/master/sources/utils/master/scheduler/tools
            $ nano initSchedulerByName.py
            
            def initSchedulerByName(
                    knownMasters: Set[Address],
                    minimumActors: int,
                    schedulerName: str,
                    basicComponent: BasicComponent,
                    isContainerMode: bool,
                    parsedArgs,
                    **kwargs) -> Union[BaseScheduler, None]:
                if schedulerName == 'OHNSGA':
                    # hidden to save space
                    pass
                elif schedulerName == 'NSGA2':
                    # hidden to save space
                    pass
                elif schedulerName == 'NSGA3':
                    # hidden to save space
                    pass
                # New Added Block
                elif schedulerName == 'RankingBased':
                    from ..policies.schedulerRankingBased import \
                        RankingBasedPolicy
                    scheduler = SchedulerRankingBased(isContainerMode=isContainerMode)
                    return scheduler
                    
                return None
            \end{lstlisting}

    \item The \textit{Master} component can be executed using the following command while the \textit{schedulerName} option shows the name of the selected scheduling policy:
            \begin{lstlisting}[language=bash,autogobble]
            $ pwd
            /home/ubuntu/fogbus2/containers/master
            $ docker-compose run --rm --name TempContainerName fogbus2-master  --containerName TempContainerName --bindIP 192.0.0.8 --schedulerName RankingBased
            \end{lstlisting} 
    
\end{enumerate}

\subsection{Evaluation Results}
To evaluate the performance of the FogBus2 framework, an integrated computing environment consisting of multiple cloud instances and edge/fog servers is prepared. Table\ref{tab:HostConfig} depicts the full configuration of servers and corresponding running FogBus2 components.

\begin{figure}
  \centering%
    \includegraphics[width=0.75\linewidth,height=4cm]{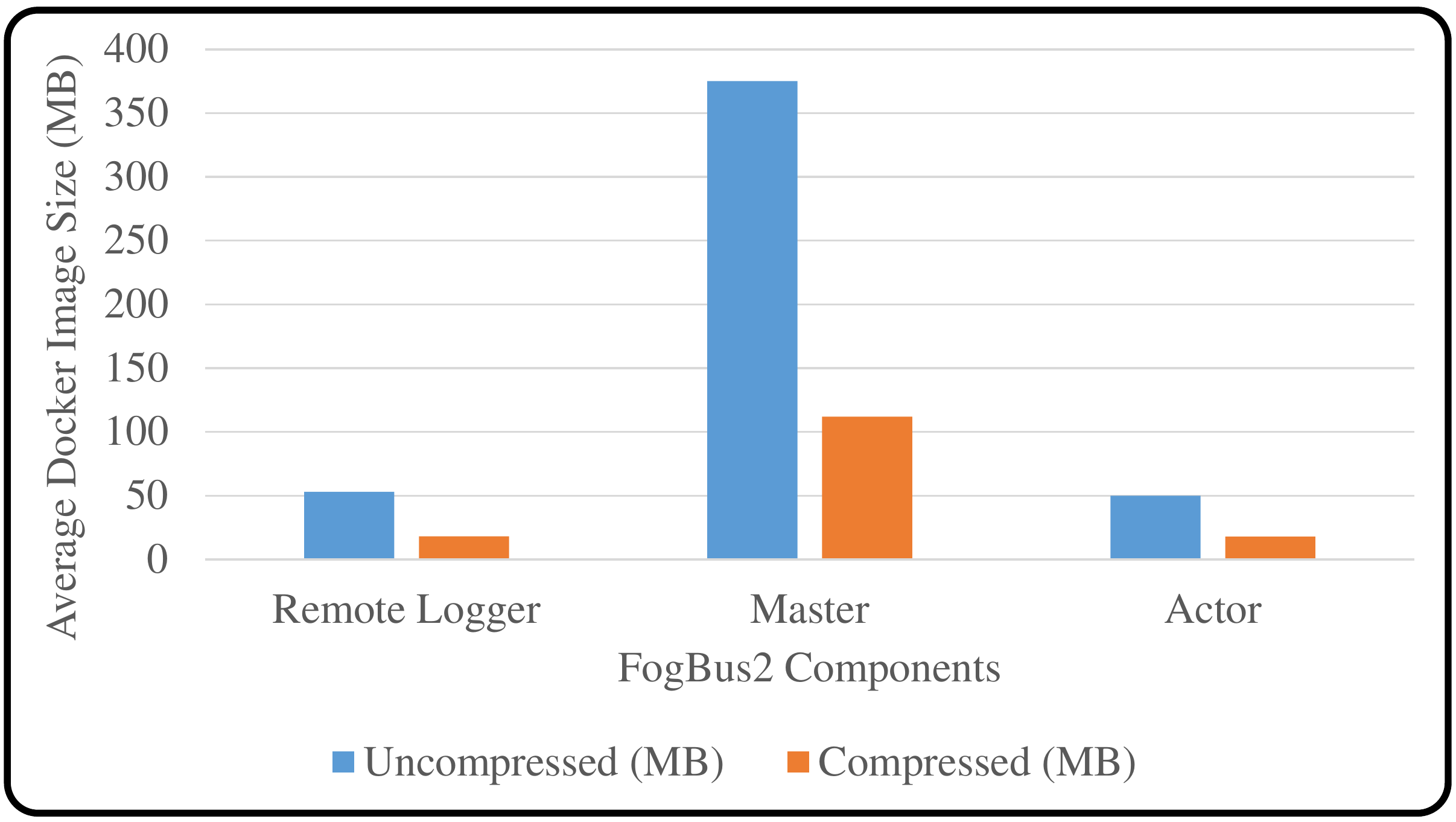}\hfill%
    \caption{Average Docker Image size of FogBus2 Components}
    \label{fig:imageSize}
    \vspace{-0.5cm}
\end{figure}
Figure~\ref{fig:imageSize} represents the average docker size of FogBus2 components in compressed and uncompressed formats. The compressed docker image size is obtained from the average size of docker images stored in the docker hub for multiple architectures, while uncompressed docker image size is obtained from the average size of extracted docker images on instances. The size of the compressed docker image shows that FogBus2 components are lightweight to be downloaded on different platforms, ranging from few megabytes to roughly 100 MB at maximum. Besides, the uncompressed docker image size proves that FogBus2 components are not resource-hungry and do not occupy the storage. The reason why the image sizes of \textit{User} and \textit{Task Executor} components are not provided is that the docker image sizes of these components heavily depend on the IoT applications.   
\begin{figure}
  \centering%
    \includegraphics[width=0.75\linewidth,height=4cm]{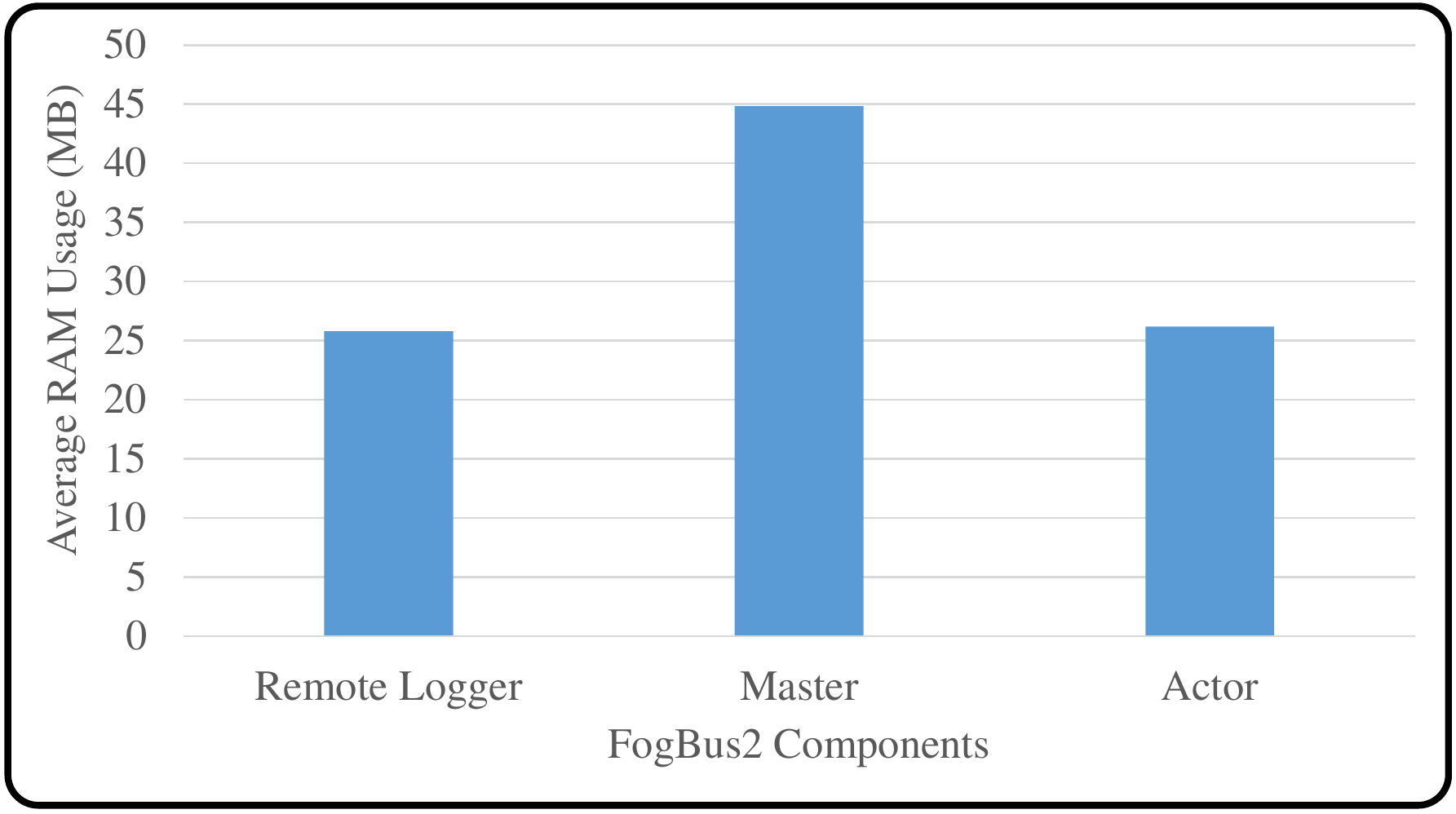}\hfill%
    \caption{Average Run-time RAM usage of FogBus2 Components}
    \label{fig:ramUsage}
    \vspace{-3mm}
\end{figure}
\par
Figure~\ref{fig:ramUsage} represents the average run-time RAM usage of FogBus2 components in different architectures. It illustrates that the average resource usage of the FogBus2 components on different architectures are low, ranging from 25 MB to 45 MB.  
\begin{figure}
  \centering%
    \includegraphics[width=0.75\linewidth,height=4cm]{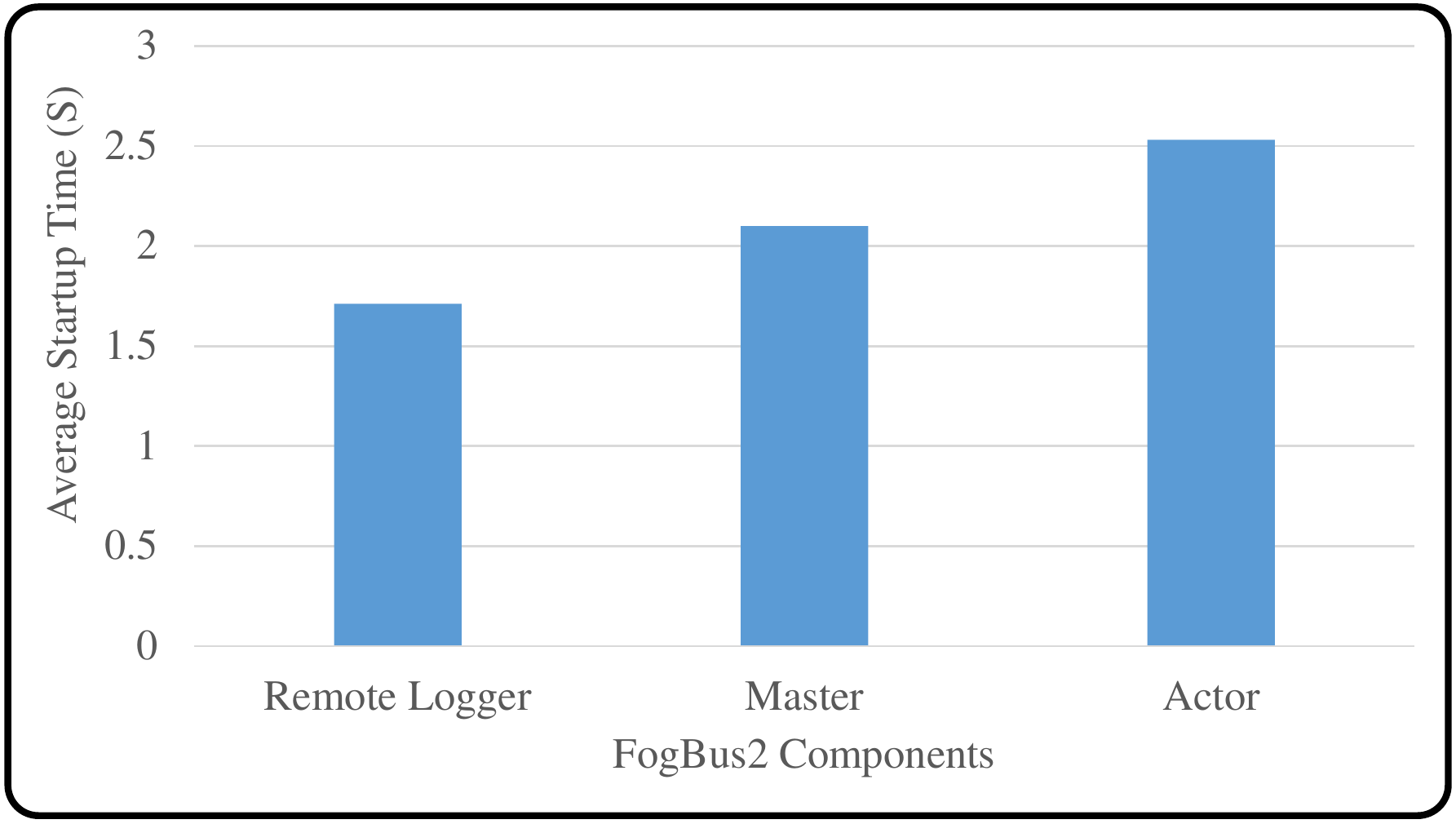}\hfill%
    \caption{Average Startup Time of FogBus2 Components}
    \label{fig:startupTime}
    \vspace{-0.2cm}
\end{figure}
Figure~\ref{fig:startupTime} demonstrates the average startup time of FogBus2 components on different architectures. It contains the amount of time required to start containers until they become in a completely functional state for serving incoming requests. Therefore, the FogBus2 framework only requires few seconds to enter into its fully functional state. It significantly helps IoT developers in the development and testing phase as they require to re-initiate the framework several times to test and debug their applications. Furthermore, in the deployment phase, it greatly helps scenarios where scalability is important.      

\begin{figure}
  \centering%
    \includegraphics[width=0.75\linewidth,height=4cm]{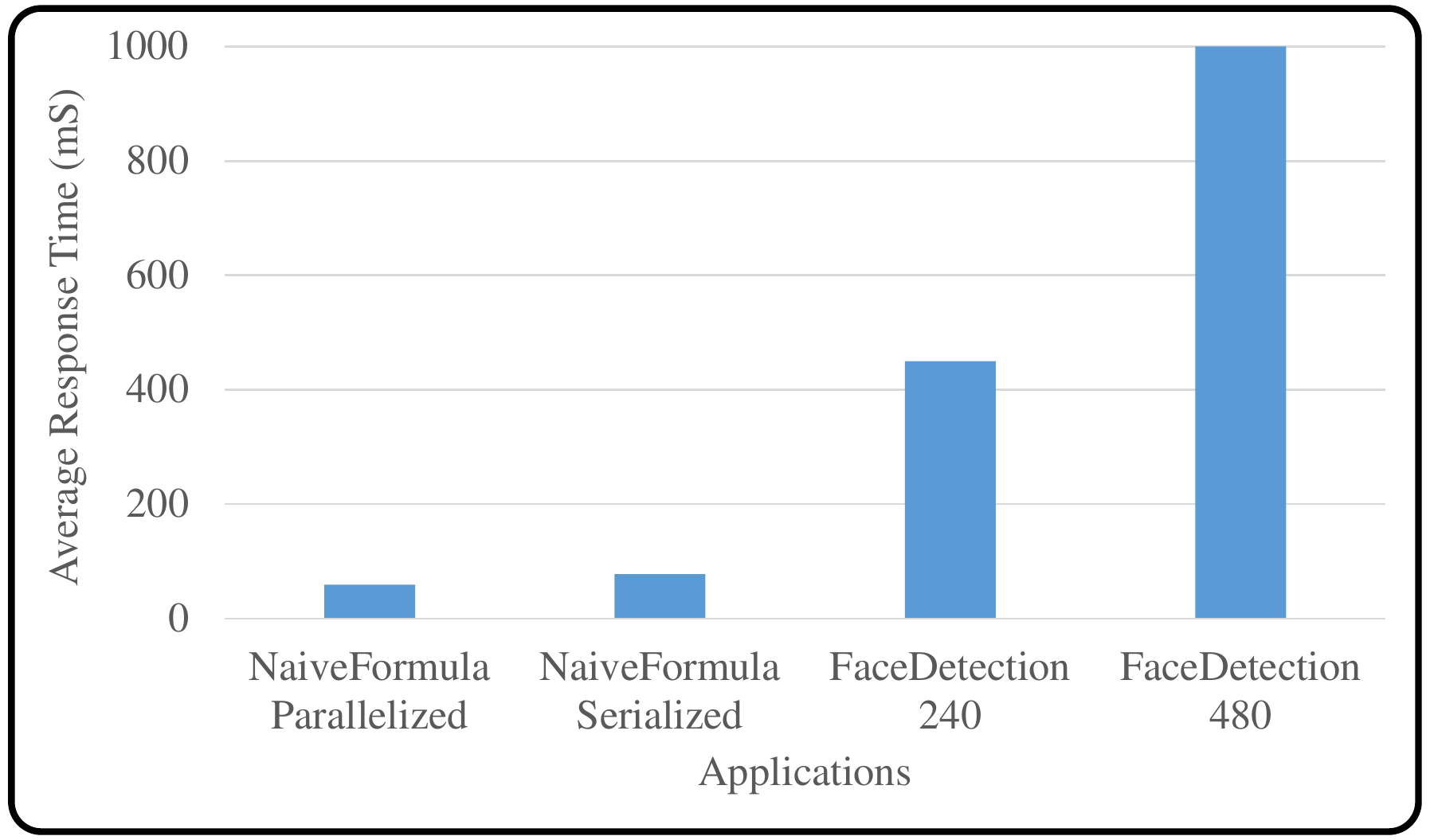}\hfill%
    \caption{Average Response Time of IoT Applications}
    \label{fig:responseTime}
\end{figure}
\par
Figure~\ref{fig:responseTime} depicts the average response time of some of the recently implemented IoT applications in FogBus2 framework.  

\section{Summary}
In this chapter, we highlighted key features of the FogBus2 framework alongside describing its main components. Besides, we described how to set up an integrated computing environment, containing multiple cloud service providers and edge devices, and establish a low-overhead communication network among all resources. Next, we provided instructions and corresponding code snippets to install and run the main framework and its integrated applications. Finally, we demonstrated how to implement and integrate new IoT applications and custom scheduling policies with this framework. 
\paragraph{\textbf{Software Availability}}
The source code of the FogBus2 framework and newly implemented IoT applications and scheduling policies are accessible from the CLOUDS Laboratory GitHub webpage: \href{https://github.com/Cloudslab/FogBus2}{https://github.com/Cloudslab/FogBus2}.

\bibliographystyle{vancouver-modified}
\bibliography{sample-vancouver}

\end{document}